# Effect of Horizontal Spacing on Natural Convection from Two Horizontally Aligned Circular Cylinders in Non-Newtonian Power-Law Fluids

**Subhasisa Rath*[1], Sukanta Kumar Dash[2]**

[1]School of Energy Science & Engineering, Indian Institute of Technology Kharagpur, 721302, India
[2]Department of Mechanical Engineering, Indian Institute of Technology Kharagpur, 721302, India

**\* Corresponding Author**
E-mail: subhasisa.rath@gmail.com
Tel.: +91-9437255862
ORCID ID: https://orcid.org/0000-0002-4202-7434

**Abstract**: Laminar natural convection from two horizontally aligned isothermal cylinders in unconfined Power-law fluids has been investigated numerically. The effect of horizontal spacing $(0 \leq S/D \leq 20)$ on both momentum and heat transfer characteristics has been delineated under the following pertinent parameters: Grashof number $(10 \leq Gr \leq 10^5)$, Prandtl number $(0.71 \leq Pr \leq 100)$, and Power-law index $(0.4 \leq n \leq 1.6)$. The heat transfer characteristics are elucidated in terms of isotherms, local Nusselt number (Nu) distributions and average Nusselt number values, whereas the flow characteristics are interpreted in terms of streamlines, pressure contours, local distribution of the pressure drag and skin-friction drag coefficients along with the total drag coefficient values. The average Nusselt number shows a positive dependence on both Gr and Pr whereas it shows an adverse dependence on Power-law index (n). Overall, shear-thinning (n<1) fluid behavior promotes the convection whereas shear-thickening (n>1) behavior impedes it with reference to a Newtonian fluid (n=1). Furthermore, owing to the formation of a chimney effect, the heat transfer increases with decrease in horizontal spacing (S/D) and reaches a maximum value corresponding to the optimal spacing whereas the heat transfer drops significantly with further decrease in S/D. Finally, a correlation for Nu has been developed, which can be useful to academic researchers and practicing engineers.

***Keywords:*** *Natural convection; Non-Newtonian fluid; Power-law fluid; Horizontal cylinder; Horizontal spacing; Correlation*

**1. Introduction:**

Natural convection heat transfer of non-Newtonian Power-law fluids have emerging potential applications in numerous process industries including polymer, food, agricultural, pharmaceutical, biomedical, biochemical, beverage and so on, where many structured fluids exhibit shear-thinning or shear-thickening behavior under pertinent industrial settings. Heating and/or cooling of such fluids are commonly encountered by natural convection. Typical examples of process engineering applications include thermal processing of food-stuffs, fruit yogurts, some packaged ready to eat foods, reheating of polymer melts or solutions, particulate suspensions containing pulps and paper fibers, chemical treatment in multiphase reactors including slurry reactors, packed and fluidized-bed reactors etc. [1-6]. Due to the temperature dependent buoyancy flow, the coupled momentum and energy equations are to be solved simultaneously which brings analytical and computational challenges in natural convection. Furthermore, the apparent viscosity of a Power-law fluid exhibits a measurable variation with shear rate in the physical domain. Thus it is exemplary to mention here that the coupled momentum and heat transfer characteristics are further accentuated subject to non-Newtonian fluids.

Owing to the pragmatic significance, extensive research on natural convection from heated cylinders of different cross-sections like circular [7-8], semi-circular [9,10], square [11], tilted square [12], and elliptical [13] in an unconfined power-law medium has been investigated numerically in recent years. In the aforementioned literature [7-13] on Power-law fluids, the heat transfer characteristics were represented in terms of local Nusselt number distributions along the surface of the heated object and the average Nusselt number. Under identical conditions, it has been found that the shear-thinning behavior of the power-law fluid enhances the heat transfer whereas shear-thickening behavior diminishes the heat transfer with reference to that of a Newtonian fluid, albeit different cross-sections and orientations of the heated object has a significant influence on the rate of heat transfer.

The preceding research on free convection is restricted to a single heated cylinder in a non-Newtonian fluid medium. Undoubtedly, appreciable physical insights into the underlying physical phenomena have been found from such research on single cylinder over a wide range of Grashof and Prandtl numbers of industrial interests [14]. However, it has been readily acknowledged that free convection from various configurations of multiple cylinders has significant practical applications. Natural convection heat transfer from multiple vertically aligned cylinders in Newtonian fluid (air) have been investigated experimentally or numerically in recent past [15-19]. It has been found from the aforementioned literature in Newtonian fluid that due to the presence of the bottom cylinder, the heat transfer from the top cylinder was reduced, exhibiting stronger interaction and preheating of the thermal plumes, which is commonly known as temperature difference imbalance in the literature. Furthermore, the ratio of Nusselt numbers of upper to lower cylinder ($Nu_u / Nu_l$) has been found to increase with an increase in the pitch to diameter ratio. Computational study on free convection from two vertically attached horizontal cylinders was done by Liu et al. [20]. Owing to the interaction of the plumes and vortex formations, the heat transfer from the individual cylinder was found to decrease compared to that of a single cylinder. The resulting heat transfer is attributed by two competing mechanisms: decreasing momentum of the flow due to preheating of the plume by the bottom cylinder (creates temperature difference imbalance) and the mixed convection behavior experienced at the top cylinder due to the plume developed by the bottom cylinder.

Notwithstanding the preceding investigations in Newtonian fluids, subsequently, this physical mechanism of free convection from multiple cylinders has been extended to non-Newtonian fluids and scant research is available in Power-law fluids. Shyam et al. [21] numerically studied the free convection from two vertically aligned circular cylinders in power-law fluids. In their study, the simulations were conducted for a wide range of pertinent conditions for both shear-thinning and shear-thickening Power-law fluids along with varying vertical spacing between the two cylinders in the range $(2 \leq (S/D) \leq 20)$. Shyam and Chhabra [22] numerically investigated the laminar natural convection from two vertically arranged square cylinders at moderate Grashof number in the range $(10 \leq Gr \leq 1000)$ in both shear-thinning as well as shear-thickening Power-law fluids $(0.4 \leq n \leq 1.8)$. Simulations were conducted by varying the inter-cylinder spacing in the range, $2 \leq L/d \leq 6$ to elucidate its effect on flow and thermal fields. It has been reported in the preceding literature [21,22] that owing to the intense interference at low L/d, the contribution of the downstream cylinder is found to be much less to the overall heat transfer and much higher to the hydrodynamic drag coefficient compared to that of the upstream cylinder.

Despite the arrangement of horizontal cylinders in a vertical array, no prior literature has been seen to be present on natural convection from a horizontal array of cylinders in non-Newtonian fluids, albeit scant literature are available in Newtonian fluid like air, [23-26]. It has been reported in the preceding investigations on Newtonian fluid that owing to the development of a chimney effect or narrow passage, the convection heat transfer increases with a decrease in the horizontal spacing between the cylinders and corresponding to the optimum spacing, it reaches a maximum value. At the optimal spacing between the cylinders, the two thermal plumes just come in contact adjacent to the cylinders and with further decrease in the horizontal spacing, the two thermal plumes tend to merge and come out as a single plume resulting in a decrease in the rate of heat transfer. Most recently, Rath and Dash [27] numerically studied the natural convection heat transfer from a stack of horizontal cylinders in air. Attached cylinders of three, six and ten in numbers were arranged in a triangular manner to form different stacks for their study in both laminar and turbulent regimes. Natural convection from a pair of two horizontally attached cylinders in air has been investigated numerically by Liu et al. [28] and it has been found that the overall heat transfer degraded with compared to that of a single cylinder due to the interaction of their thermal plumes. In addition, two recirculation zones were seen to form in their work when the two plumes merged and the size of these vortices was found to grow with the increase in Rayleigh number.

To the best of our knowledge, the literature on non-Newtonian fluids lacks a comprehensive study on natural convection from a pair of two horizontally aligned circular cylinders. Hence, the present work is motivated towards the computational study on laminar free convection from two horizontally aligned cylinders in order to address the effect of horizontal spacing between the two cylinders and the role of power-law rheology on both momentum and heat transfer characteristics. The applications of such geometrical configurations can be found in polymer processing industries where multiple heated cylinders of different shapes are used for natural convection heating of several non-Newtonian fluids. The reported results herein include the streamlines, isotherm contours, and pressure contours in close proximity of the cylinders, local and average values of Nusselt number, distribution of local pressure drag and friction drag coefficients along with the total drag coefficients. Finally, the paper is concluded with the development of a new Nusselt number correlation as a function of horizontal spacing, Grashof number, Prandtl number, and Power-law index which can be beneficial to practicing engineers and academic researchers.

## 2. Problem Formulation

### *2.1. Physical description*

To perform the numerical investigations of external buoyancy-driven convection from two horizontally aligned heated cylinders, it is customary to immerse the cylinders in an extensive unconfined fluid domain of sufficiently large size. Figure 1 shows the schematic representation of the physical problem, where a pair of two equi-diameter horizontal cylinders are placed horizontally in a two-dimensional computational domain. The cylinders are considered to be very long in the z-direction. Hence, 2D simulation supports it, as the variation of any field property would be negligible in the z-direction. Gravity is set up in the negative Y-direction. For the Cartesian coordinate system, the origin is set exactly at the center of the domain. The heated cylinders are maintained at a fixed temperature of $T_w$ which are immersed in a surrounding quiescent non-Newtonian power-law fluid maintained at temperature $T_\infty$ ($T_w > T_\infty$). The presence of temperature gradient leads to thermal expansion of the fluid in the domain and as a result, the fluid near the heated cylinders will be lighter compared to the fluid far away from the cylinders. Owing to the pressure gradient, an upward flow or buoyancy will set up adjacent to the cylinders which in turn, results in natural convection of fluid in the domain. The objective is to assess the flow and thermal field around the attached cylinders and hence predict the Nusselt number and drag coefficients for it.

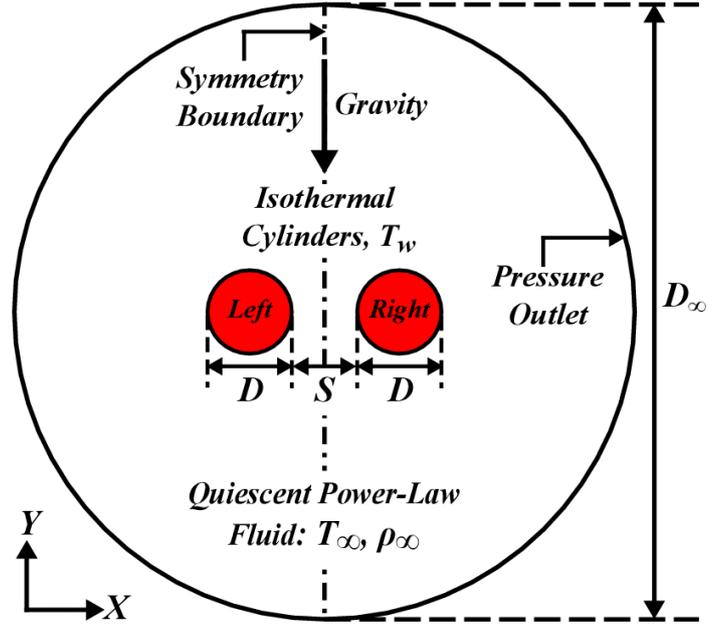

**Figure 1.** Schematic representation of the physical domain

## 2.2. Mathematical Modeling

*2.2.1. Assumptions:* The flow is assumed to be steady, laminar, buoyancy-driven, incompressible, two-dimensional, and symmetric about the vertical centreline. The shear-dependent viscosity of the fluid has been captured by using the non-Newtonian power-law model. The thermo-physical properties (except density) of the fluid are assumed to be independent of temperature and evaluated at mean film temperature ($T_m$). The temperature dependent density is taken into consideration by employing Boussinesq approximation as Eq.(1):

$$(\rho_\infty - \rho) \approx \rho \beta (T - T_\infty) \tag{1}$$

It is worthwhile to mention here that to justify the assumption of the use of constant thermo-physical properties at mean film temperature ($T_m$) for non-Newtonian fluids, the maximum temperature difference $(\Delta T = T_w - T_\infty)$ should never exceed 5K in the system [29-31]. The maximum temperature difference for all the simulations in the present study is set to $\Delta T = 1K$ which is quite reasonable to assume the thermal expansion coefficient $(\beta)$ as a constant value as expressed in Eq.(2):

$$\beta = -\frac{1}{\rho}\left(\frac{\partial \rho}{\partial T}\right)_P = \frac{1}{T_m} \tag{2}$$

Furthermore, in the present study, the radiative heat transfer is assumed to be negligible since temperature being very small and the viscous dissipation effect is also neglected owing to the maximum value of Brinkman number, Br << 1.

*2.2.2. Governing Equations:* With reference to the aforementioned suitable assumptions, the coupled flow and thermal fields can be written with the governing conservation equations (continuity, momentum, and thermal energy) in their non-dimensional form as follows:

*Continuity:*
$$\frac{\partial U_X}{\partial X} + \frac{\partial U_Y}{\partial Y} = 0 \tag{3}$$

*X- Momentum:*
$$\frac{\partial (U_X U_X)}{\partial X} + \frac{\partial (U_X U_Y)}{\partial Y} = -\frac{\partial P}{\partial X} + \frac{1}{\sqrt{Gr}} \left( \frac{\partial \tau_{XX}}{\partial X} + \frac{\partial \tau_{YX}}{\partial Y} \right) \tag{4}$$

*Y- Momentum:*
$$\frac{\partial (U_Y U_X)}{\partial X} + \frac{\partial (U_Y U_Y)}{\partial Y} = -\frac{\partial P}{\partial Y} + \varphi + \frac{1}{\sqrt{Gr}} \left( \frac{\partial \tau_{XY}}{\partial X} + \frac{\partial \tau_{YY}}{\partial Y} \right) \tag{5}$$

*Thermal Energy:*
$$\frac{\partial (U_X \varphi)}{\partial X} + \frac{\partial (U_Y \varphi)}{\partial Y} = \frac{1}{Gr^{1/(n+1)} Pr} \left( \frac{\partial^2 \varphi}{\partial X^2} + \frac{\partial^2 \varphi}{\partial Y^2} \right) \tag{6}$$

For an incompressible fluid, the deviatoric stress tensor ($\tau_{ij}$) is directly associated with the rate of deformation tensor ($\varepsilon_{ij}$), as given by Bird at al. [32]:

$$\tau_{ij} = 2\eta \varepsilon_{ij} \tag{7}$$

The rate of deformation tensor or strain rate tensor ($\varepsilon_{ij}$) is in turn associated with the velocity gradient field [32] and can be written as:

$$\varepsilon_{ij} = \frac{1}{2} \left[ \frac{\partial U_i}{\partial j} + \frac{\partial U_j}{\partial i} \right] \tag{8}$$

The non-Newtonian viscosity $(\eta)$ for a power-law fluid is interlinked with the second invariant of the strain rate tensor as given by Bird at al [32], which in turn defined as:

$$\eta = m \left( \frac{I_2}{2} \right)^{(n-1)/2} \tag{9}$$

Where the second invariant $(I_2)$ of the strain rate tensor is defined as:

$$I_2 = \sum_i \sum_j \varepsilon_{ij} . \varepsilon_{ji} \tag{10}$$

All the non-dimensional variables appearing in Eqs. (3-6) are defined as follows:

$$X = \frac{x}{L}, Y = \frac{y}{L}, U_X = \frac{u_x}{u_c}, U_Y = \frac{u_y}{u_c}, P = \frac{p}{\rho_\infty u_c^2}, \varphi = \frac{T - T_\infty}{T_w - T_\infty} \tag{11}$$

Where, L is the characteristic length scale (defined as; L = D) and $u_c$ is the characteristic reference velocity (defined as: $u_c = \sqrt{g\beta \Delta T L}$) used for the present simulations.

In the present study, the apparent viscosity of a non-Newtonian fluid is approximated by using the Ostwald–de Waele power-law viscosity model as follows:

$$\eta = m \, \dot{\gamma}^{n-1} \quad (12)$$

*2.2.3. Boundary Conditions:* To solve the governing conservation Eqs. (3-6), some suitable physically realistic boundary conditions are imposed on the computational domain, which can be summarized in mathematical form as follows.

*On the cylinder surface:* The surface of the cylinders are assigned as wall boundary with isothermal condition. No slip and no penetration boundary conditions are imposed on these walls, i.e.:

$$U_X = U_Y = 0 \text{ and } \varphi = 1 \quad (13)$$

*At the outer boundary:* The outer surface of the flow domain is designated as pressure outlet boundary where pressure is set to be the atmospheric pressure (zero static gauge pressure). Pressure outlet boundary would allow the fluid to either go out of the domain or come into the domain depending upon the inside pressure condition in the domain. The temperature of any back-flow would be same as the surroundings quiescent fluid temperature, i.e.:

$$P = P_\infty = 0 \text{ (gauge)}; \; \varphi = 0 \text{ (back-flow temperature)} \quad (14)$$

*At the vertical axis (X = 0):* Owing to the appearance of steady and symmetric flow about the central vertical axis (at $X = 0$) of the computational domain, the computations are conducted only with a half computational domain $(X \geq 0)$. Hence, a planar symmetry boundary is imposed at this vertical centerline.

$$U_X = 0, \frac{\partial U_Y}{\partial X} = 0 \text{ and } \frac{\partial \varphi}{\partial X} = 0 \quad (15)$$

### 2.3. Heat transfer and fluid parameters

The numerical solution represents the flow domain in terms of some field variables, i.e., velocity, temperature, and pressure. Evidently, in the present study, the coupled momentum and thermal fields are influenced by four dimensionless parameters, i.e., Grashof number (Gr), Prandtl number (Pr), power-law index (n), and horizontal spacing (S/D) which are defined as follows for non-Newtonian power-law fluids.

*Grashof number:*

$$Gr = \frac{\rho^2 D^{n+2} (g\beta\Delta T)^{2-n}}{m^2} \quad (16)$$

*Prandtl number:*

$$Pr = \left(\frac{\rho C_p}{K}\right)\left(\frac{m}{\rho}\right)^{\left(\frac{2}{1+n}\right)} D^{\left(\frac{1-n}{1+n}\right)} (g\beta\Delta T D)^{\frac{3(n-1)}{2(n+1)}} \quad (17)$$

Moreover, the power-law index, in its own right, is a non-dimensional parameter but owing to its appearance in both Grashof and Prandtl numbers, it increases the level of difficulties in particularising its role on the numerical results. In the present numerical study, the horizontal spacing, power-law index, Grashof number, and Prandtl number are considered for a wide range of pertinent industrial conditions such as; $(0 \leq S/D \leq 20)$, $(0.4 \leq n \leq 1.6)$, $(10 \leq Gr \leq 10^5)$, and $(0.71 \leq Pr \leq 100)$, respectively. Where, n<1 indicates the shear-thinning fluid behavior, n > 1 indicates the shear-thickening fluid behavior and n=1 corresponds to the generalized Newtonian fluid behavior.

In the present study, the numerical results are summarized in §4 of this paper, in terms of streamlines, temperature and pressure contours, Nusselt numbers, and drag coefficients as a function of horizontal spacing, Grashof number, Prandtl number, and power-law index in the aforementioned ranges. Thus, it is prudent to describe some of these parameters here.

*Nusselt number (Nu):* The convective heat transfer is associated with the non-dimensional parameter called Nusselt number which is the ratio of convective to conductive heat transfer at the surface. The average Nusselt number is numerically calculated by taking the area weighted average or integrating the values of local Nusselt numbers over the cylinder surface as written below:

$$Nu = \int_s Nu_l \, ds \qquad (18)$$

Where the local Nusselt number value is calculated as follows:

$$Nu_l = \frac{h_l D}{K} = -\left(\frac{\partial \varphi}{\partial \hat{n}_s}\right) \qquad (19)$$

*Drag coefficient ($C_d$):* It is associated with the total drag force, which is the resultant force on the cylinders in the direction of buoyancy due to the gradients of velocity appearing in the vicinity of the cylinders. The resulting drag force is the sum of the normal pressure force and shearing force. Hence, the total drag coefficient ($C_d$) has two components such as pressure drag coefficient ($C_{pr}$) and skin friction drag coefficient ($C_{sf}$), which are due to pressure difference in the direction of flow and shear stress, respectively.

$$C_d = C_{pr} + C_{sf} \qquad (20)$$

Where,
$$C_{pr} = \frac{F_{d,pr}}{1/2 \rho_\infty u_c^2 D} = \int_s C_{pr,\theta} \, \hat{n}_y ds \qquad (21)$$

The local pressure drag coefficient:
$$C_{pr,\theta} = \frac{P_\theta - P_\infty}{1/2 \rho_\infty u_c^2} \qquad (22)$$

and
$$C_{sf} = \frac{F_{d,f}}{1/2 \rho_\infty u_c^2 D} = \int_s C_{sf,\theta} \, \hat{n} \, ds \qquad (23)$$

The local skin friction coefficient: $C_{sf,\theta} = \dfrac{\tau_\theta}{1/2\rho_\infty u_c^2} = \dfrac{2^{1-n}}{\sqrt{Gr}}\left(\dfrac{\partial U_\theta}{\partial \hat{n}}\right)_{wall}$  (24)

## 3. Numerical Procedure

The aforementioned governing Eqs.(3-6) along with the imposed boundary conditions (Eqs.(13-15)) were solved numerically by using a finite volume method (FVM) based algebraic multi-grid (AMG) solver of the commercial software package ANSYS-Fluent-v15. For coupling the pressure-velocity terms, the Semi-Implicit Method for Pressure-Linked Equations (SIMPLE) algorithm has been implemented since it was found to be the most stable one. PRESTO (PREssure STaggering Option) scheme was employed for pressure discretization which calculates the pressure at the face center and avoids the interpolation errors and hence, gives more accurate results. Second Order Upwind (SOU) scheme was applied for discretization of the convective terms. Furthermore, for a smooth convergence, the scaled residuals or the convergence criterions were set to a very low value (unlike as for Newtonian fluids) such as: for energy equation, it was set to $10^{-13}$ and for continuity and momentum equations, it has been set to $10^{-8}$. These criterion values are more adequate for the present study since the monitor of Nusselt number and drag coefficient was consistent up to five significant digits under these criterions. In addition, the non-Newtonian power-law model and Boussinesq approximation were used for capturing the shear-dependent viscosity and temperature dependent density, respectively. It is customary to mention here that the use of any specific values of thermo-physical properties has no importance since the results are reported here in their non-dimensional form. The under-relaxation factors used for the present simulations are given in Table 1.

**Table 1.** Under-relaxation factors:

| Pressure | Density | Body forces | Momentum | Energy |
|---|---|---|---|---|
| 0.5 | 1 | 1 | 0.4 | 1 |

### *3.1. Choice of computational domain and grid distributions:*

To conduct the numerical simulations for a purely unconfined flow problem is quite impossible. Hence, a sufficiently large cylindrical fluid domain of diameter $D_\infty$ has been considered for the present study as shown in Figure 1. The flow is expected to be steady and symmetric flow about the vertical centreline (X = 0) of the computational domain. Hence, symmetry condition has been taken in the present simulations and computation has been performed for half computational domain only (X ≥ 0) as shown in Figure 2 which would minimize the computational cells and resources to a great extent. It is a common intuition that near the heated cylinders the temperature and velocity gradients will vary appreciably whereas, the gradient of field properties will vary moderately far away from the cylinders. Hence, the variation of temperature and velocity distributions were captured by providing very fine cells over the cylinders using inflation technique (first layer thickness = 0.01D and growth rate = 1.2) for first 20 cells from the cylinder wall. In contrast, it is just an unnecessary computational effort to use fine mesh in a region far away from the cylinders. Keeping this in mind, relatively courser quadrilateral mapped cells are provided far away from the cylinders as shown in Figure 2.

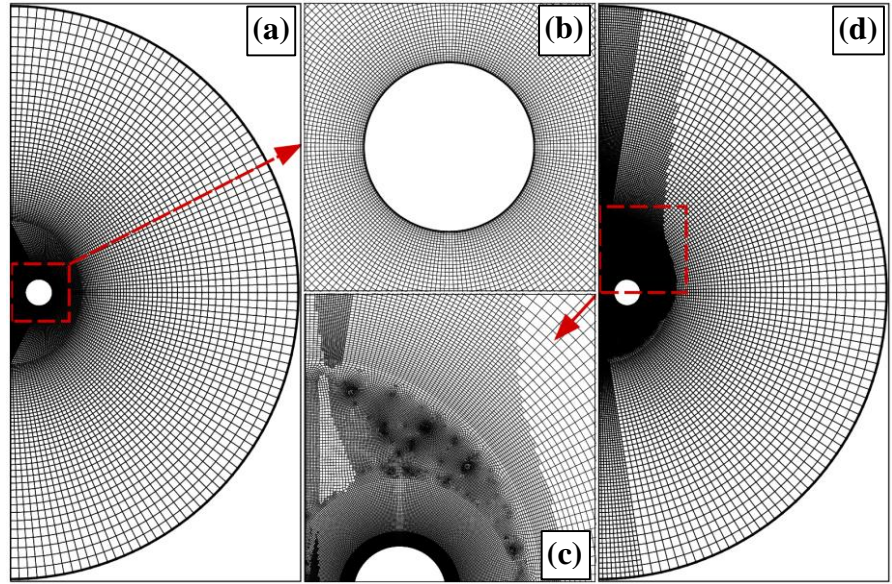

**Figure 2.** Schematic representation of grid distributions in half computational domain; (a) initial grids (without adaption), (b) blown-up view near the cylinders (without adaption), (c) blown-up view near the cylinders (with adaption) and, (d) grids with adaption.

### *3.2. Domain independence test:*

The size of the computational domain would affect the numerical results to some extent. Hence, a domain-independent study is carried out such that it would not affect the flow field and also keep the requisite computational resources at an optimum level at the same time. For the present study, domain-independent test was carried out by varying the domain size, $D_\infty$ in the range spanning from 50D to 800D at a lowest value of Prandtl number (Pr = 0.71) where the boundary layers are expected to be quite thick and at two extreme values of power-law index (n = 0.4 and n = 1.6) as shown in Table 2. Domain-independent study has been done for individual Grashof numbers and the domain size is fixed to 600D, 400D and 200D for Gr = 10, Gr = $10^3$ and Gr = $10^5$, respectively which are found to be sufficient for the present study since, the relative variation in average Nusselt number beyond this size were found to be less than 1%.

**Table 2.** Domain independence test (half-domain): Variation of average Nusselt number with domain size at Pr = 0.71 for S/D = 1

| Domain, $D_\infty$ | Gr = 10 | | Gr = $10^3$ | | Gr = $10^5$ | |
|---|---|---|---|---|---|---|
| | n = 0.4 | n = 1.6 | n = 0.4 | n = 1.6 | n = 0.4 | n = 1.6 |
| 50 D | 0.9365 | 0.7392 | 4.7473 | 1.9592 | 20.4539 | 4.1846 |
| 100 D | 0.9989 | 0.7938 | 4.9825 | 2.0579 | 21.0545 | 4.3122 |
| 200 D | 1.0429 | 0.8283 | 5.1391 | 2.1191 | **21.3714** | **4.4311** |
| 400 D | 1.0706 | 0.8503 | **5.2456** | **2.1467** | 21.4265 | 4.4709 |
| 600 D | **1.0867** | **0.8654** | 5.2620 | 2.1559 | 21.4415 | 4.4962 |
| 800 D | 1.0896 | 0.8681 | 5.2708 | 2.1587 | --- | --- |

## *3.3. Grid independence test:*

Grid independence study has also been carried out to ensure the numerical results remain to be independent of the computational cell size. In the present study, grid independence test was carried out by using adaption technique at the highest Grashof number (Gr = $10^5$) and Prandtl number (Pr = 100) where the boundary layers are expected to be quite thin over the cylinders and at two extreme values of power-law index (n = 0.4 and n = 1.6) as shown in Table 3. After getting a converged solution using the initial mesh (no adaption), boundary adaption of 10 number of cells has been implemented on the cylinder walls and symmetry boundaries and after which successive gradient adaptions of velocity (refined value = 10% of maximum velocity magnitude) were used to get a grid independent solution. The schematics of the adapted grids are shown in Figure 2. From Table 3 it can be seen that the third adaption is found to be grid independent for the present computations since the relative variation in average Nusselt number with its succeeding adaption is less than 1%. Hence, the third adaption is incorporated in all the simulations in the present study before reporting any result.

**Table 3.** Grid independence test (half-domain): Variation of average Nusselt number with grid size at Gr = $10^5$ and Pr = 100 for S/D = 1

| Adaption | No. of cells | n = 0.4 Nu | n = 1.6 Nu |
|---|---|---|---|
| No Adaption | 42854 | 79.4735 | 21.9775 |
| 1st (Boundary) | 58252 | 73.0145 | 20.2957 |
| 2nd (Gradient) | 81694 | 69.3939 | 19.4815 |
| **3rd (Gradient)** | **108055** | **67.8016** | **19.0757** |
| 4th (Gradient) | 164688 | 67.2867 | 18.9881 |

## 4. Results and Discussions

### *4.1. Validation of the numerical methodology*

The present numerical methodology has been validated with the numerical results of Prhashanna and Chhabra [7] for an unconfined natural convection heat transfer from a single horizontal cylinder in non-Newtonian power-law fluids as shown in Figure 3. The local Nusselt number distribution over the cylinder is shown in Figure 3(a) at $Gr = 10^3$ and n = 1.8 for two different Prandtl number values of 7 and 50. Furthermore, the average surface Nusselt numbers with respect to the power-law index, n are shown in Figure 3(b) and Figure 3(c) for Gr = 10 and $10^3$ respectively, for three different Pr of 7, 20, and 50. It can be seen from Figure 3 that the present simulation results for a single cylinder show a very good agreement with the numerical results of Prhashanna and Chhabra [7]. In addition, the present numerical results for a pair of attached cylinders (S/D =0) has been validated with the numerical work of Liu et al. [28] in air for Pr=0.71 and n=1. Figure 4 shows the comparison of average Nusselt number between the present work and the numerical work of Liu et al. [28] for horizontally attached cylinders where the relative errors were found to be less than 3%. From the comparisons shown in Figure 3 and Figure 4, we could get enough confidence that our computation is accurate and reliable, and eventually may give us satisfactory results for natural convection heat transfer in non-Newtonian power-law fluids

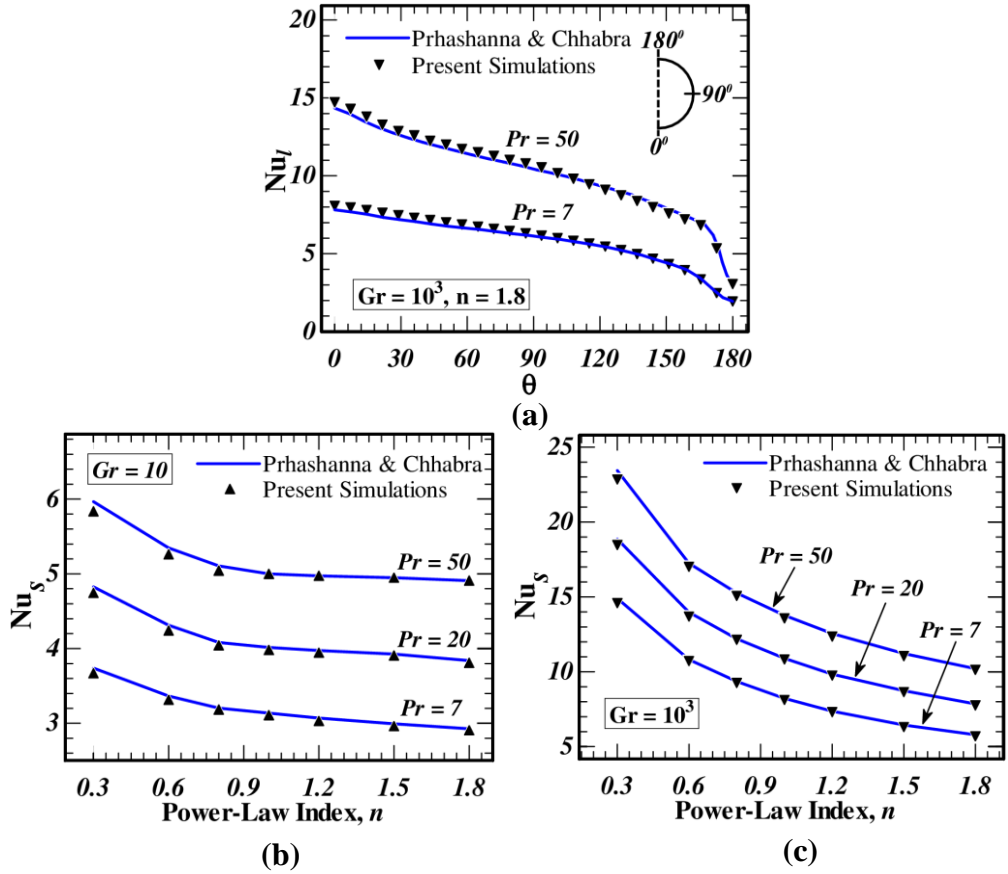

**Figure 3.** Comparison of local and average Nusselt numbers of a single cylinder with the numerical results of Prhashanna & Chhabra [7], (a) distribution of local Nusselt number over the cylinder surface, (b) average Nusselt number as a function of power-law index at Gr = 10 and (c) average Nusselt number as a function of power-law index (n) at Gr = $10^3$

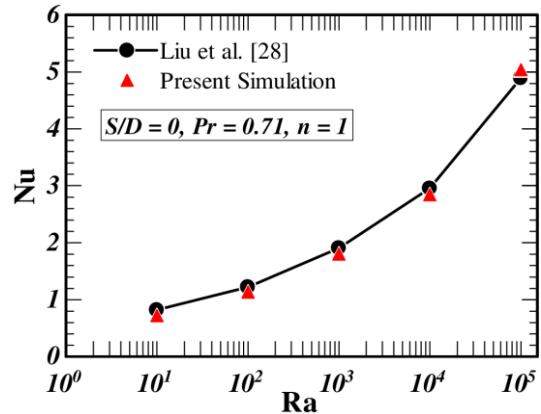

**Figure 4.** Comparison of average Nusselt number of the present work with the numerical study of Liu at al. [28] for S/D = 0, Pr = 0.71 and n = 1

## *4.2. Streamlines and pressure contours*

The flow fields are visualized in terms of streamlines and pressure contours in the close proximity of the heated cylinders. Figure 5 and Figure 6 show the representative streamline patterns (left half) and normalized pressure contours (right half) at Gr = 10 and $10^3$, respectively. Owing to the thermal expansion or density gradient, the fluid is dragged towards the heated cylinders from the pressure outlet boundary and a buoyancy-driven flow is set up thereby resulting in the formation of a plume adjacent to and above the cylinders. At low Grashof number (Gr = 10), the strength of the buoyancy is so weak that the flow due to buoyancy levitates from beneath the cylinders and flows vertically upward by sliding over the cylinder surfaces whereas, the strength of the buoyancy increases with increase in the Grashof number and a significant amount of fluid is drawn from the sides of the cylinders.

Qualitatively, in the underlying physics of natural convection, the maximum value of the effective shear-rate is expected near the isothermal cylinders which gradually diminishes away from the cylinders and vanishes to zero where the power-law fluid is almost stationary. Thus, the shear-dependent viscosity is minimum near the cylinders for a shear-thinning fluid which progressively increases and reaches to a maximum value away from the cylinders. Thus, an imaginary cavity like fluid structure is formed, which is virtually bounded by the surrounding highly viscous fluid. Hence, the shear-thinning fluid always tries to stabilize the flow and consequently eliminates the possibility of flow separations and vortex formations. In contrast, a completely contradictory behavior can be seen in shear-thickening fluids.

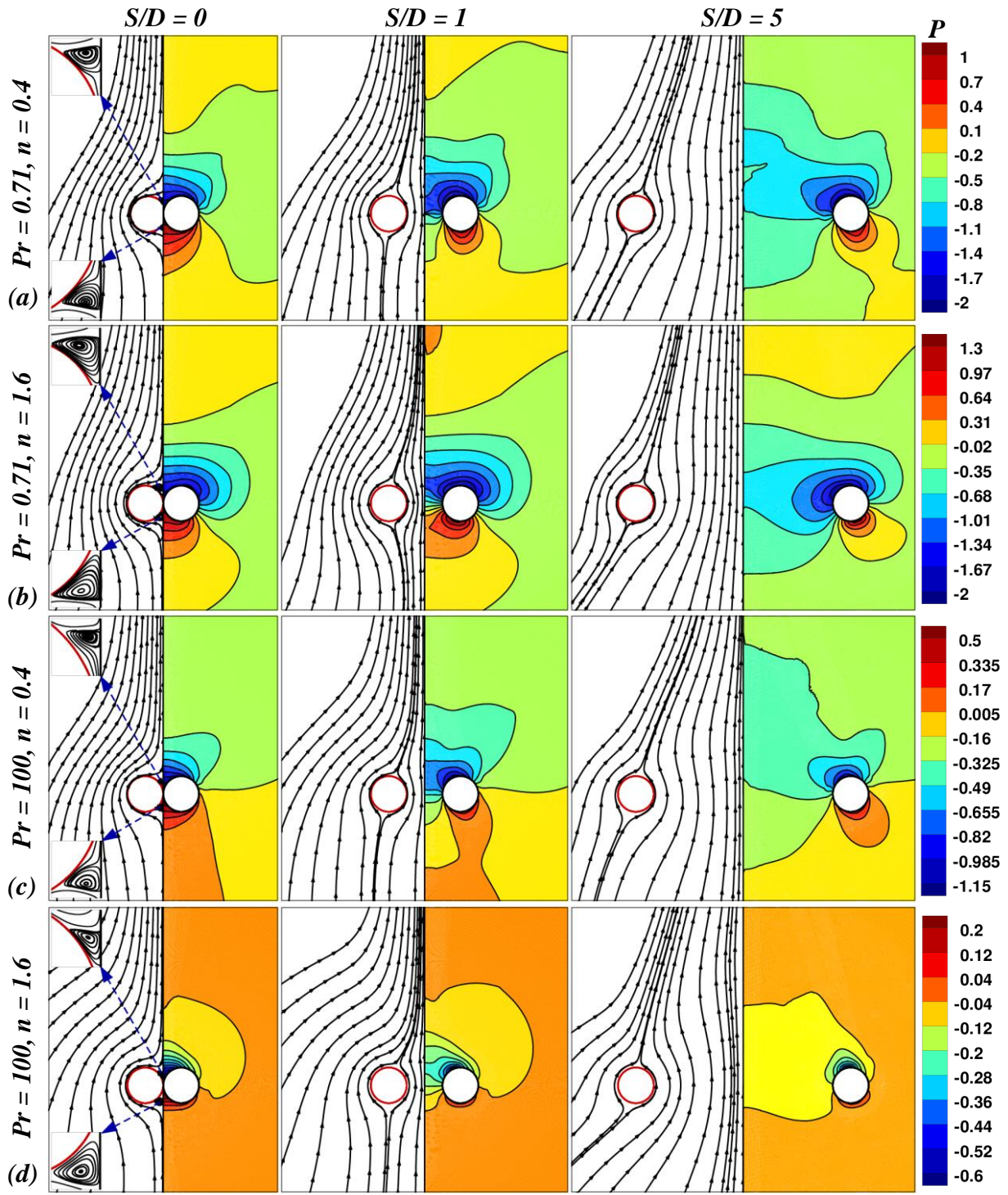

**Figure 5.** Streamlines (left side) & non-dimensional pressure contour (right side) at Gr = 10

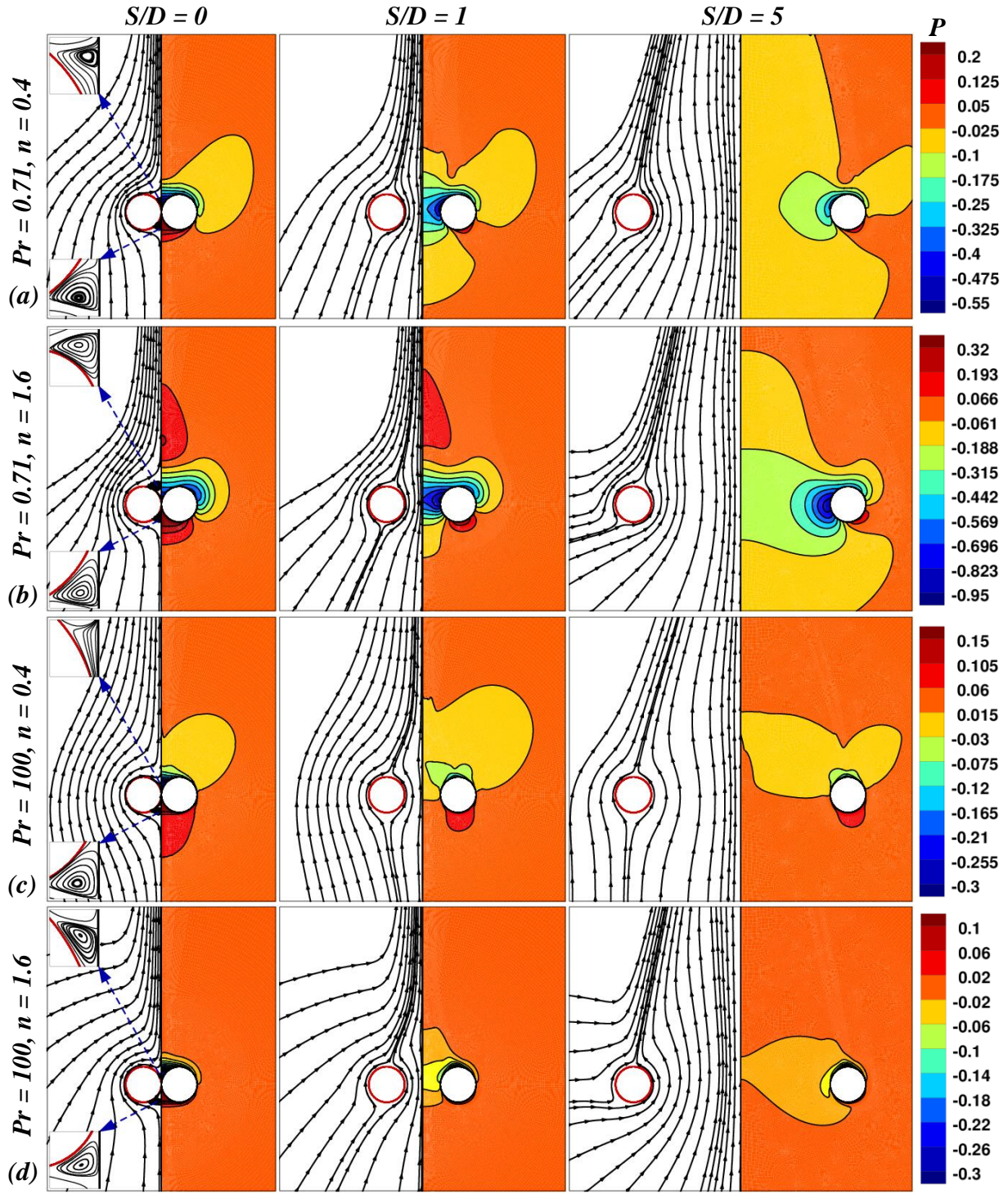

**Figure 6.** Streamlines (left side) & non-dimensional pressure contour (right side) at $Gr = 10^3$

At S/D = 0 (attached cylinders), it is worthwhile to notice in Figure 5 and Figure 6 that two recirculation zones are developed at above (back region) and below (front region) the contact point of the two cylinders. The contact point of the two cylinders and the stagnation point or the flow separation point on the surface of the cylinders are accountable for the vortex formations. The size of these vortices progressively increases with increasing value of the Grashof number and power-law index whereas the size progressively reduces with increase in Prandtl number. The precise justifications for this observation can be explained qualitatively as: at low Grashof number, fluid rises from beneath the two cylinders and the stagnation point of the front vortex moves towards the contact point. Hence, small size vortices are seen to form whereas, when Grashof number increases, more fluid entrains from the sides and the front stagnation point gradually moves downstream and the back stagnation point gradually moves upstream. Hence, the flow separation advances and gradually the size of the vortices increases. Owing to the thinning of boundary layers and increase in velocity gradient, the flow separation delays and the size of the vortices reduces with an increase in Prandtl number. As discussed in the preceding paragraph, the shear-thinning (n < 1) fluid stabilizes the flow, hence with increase in the power-law index, the size of the vortices gradually increases. For S/D > 0, flow is allowed to pass in the gap between the cylinders and as a result, it eliminates the chances of flow separations and vortex formations. It can be seen from the streamline patterns that owing to the development of a chimney effect at low S/D (approximately $0.5 \leq S/D \leq 2$), the momentum of the flow in the narrow passage between the cylinders is very high compared to that at high S/D (S/D >2). Hence, at the optimal horizontal spacing, flow moves faster over the surface of the cylinder on the symmetry side, whereas a smooth movement of the flow can be seen on the other side of the cylinders.

The pressure contours can reveal some of the physical insights into the flow field. Hence, non-dimensional pressure contours are shown in Figure 5 and Figure 6 to elucidate the effect of horizontal spacing on pressure distribution in the computational domain for some pertinent conditions. With increase in Grashof number and Prandtl number, the range of pressure variation gradually decreases due to increase in the momentum of the flow over the cylinder. In addition, with increase in the power-law index from shear-thinning to shear-thickening fluid behavior, the range of pressure variation increases at low Prandtl number, whereas it decreases for high Prandtl number. Pressure is expected to be maximum at a region where the velocity of the flow is almost zero and vice-versa. Hence, for the case of attached cylinders (S/D=0), the high pressure and low-pressure zones are formed just below and above the contact point of the two cylinders, respectively. With increase in the horizontal spacing, the high-pressure zone gradually moves downstream and the low-pressure zone gradually moves upstream. Furthermore, it can be clearly seen from the pressure contours that with increase in the horizontal spacing, the high-pressure zone progressively shrinks and the low-pressure zone grows. Hence, one can anticipate a stronger momentum of the flow and relatively higher heat transfer at S/D > 0 compared to that of the no spacing case (S/D =0).

These streamline patterns and pressure contours are completely different from that of a single cylinder where the vortex formations and flow interactions are rarely seen. Hence, the momentum and heat transfer characteristics are expected to be different prominently for the two horizontally aligned cylinders.

### *4.3. Effect of S/D on Thermal field*

The thermal fields are visualized in terms of normalized temperature contours (isotherms) around the heated cylinders as shown in Figure 7 for some combinations of Gr = 10 & $10^3$, Pr = 0.71 & 100 and n = 0.2 (shear-thinning) & 1.8 (shear-thickening). Newtonian behavior (n = 1) of the fluid is not shown here for the sake of brevity. Irrespective of the power-law index, the thermal boundary layers are quite thick at a low value of Gr and Pr and gradually becoming thinner with increase in the strength of the buoyancy or with an increase in Gr or Pr. Furthermore, it can be anticipated from the preceding discussion that the average Nusselt number would show a positive dependence on both Gr and Pr, albeit this dependence of heat transfer is more prominent on Gr compared to Pr, as evident in Figure 7. Moreover, under identical conditions, the boundary layers are found to be marginally thin in shear-thinning and thick in shear-thickening fluid behaviors with reference to the Newtonian fluids. Thus, one can anticipate an augmentation in heat transfer in shear-thinning fluids and depletion in heat transfer in shear-thickening fluids.

The effect of horizontal spacing plays a vital role in the thermal fields. At S/D = 0, fluid is not allowed to pass between the cylinders, therefore a thick thermal boundary layer is seen just above the attached cylinders and hence, the heat transfer rate is expected to be lower at this hot-spot. In contrast, at high S/D, two separate thermal plumes are seen to form around the heated cylinders which may or may not merge with each other at the top depending on the Gr and Pr as shown in Figure 7. With decrease in S/D, owing to the development of chimney effect (narrow passage between the cylinders), the momentum of the buoyancy increases and as a result, the thermal boundary layer gradually becoming thinner. Hence, the convection heat transfer increases with a decrease in the horizontal spacing and corresponding to the optimal spacing, the heat transfer is expected to reach a maximum value. At the optimal spacing between the cylinders, the two thermal plumes just come in contact adjacent to the cylinders and with further decrease in the horizontal spacing, the two thermal plumes tend to merge and come out as a single plume resulting in a thicker thermal boundary layer which decrease the rate of heat transfer.

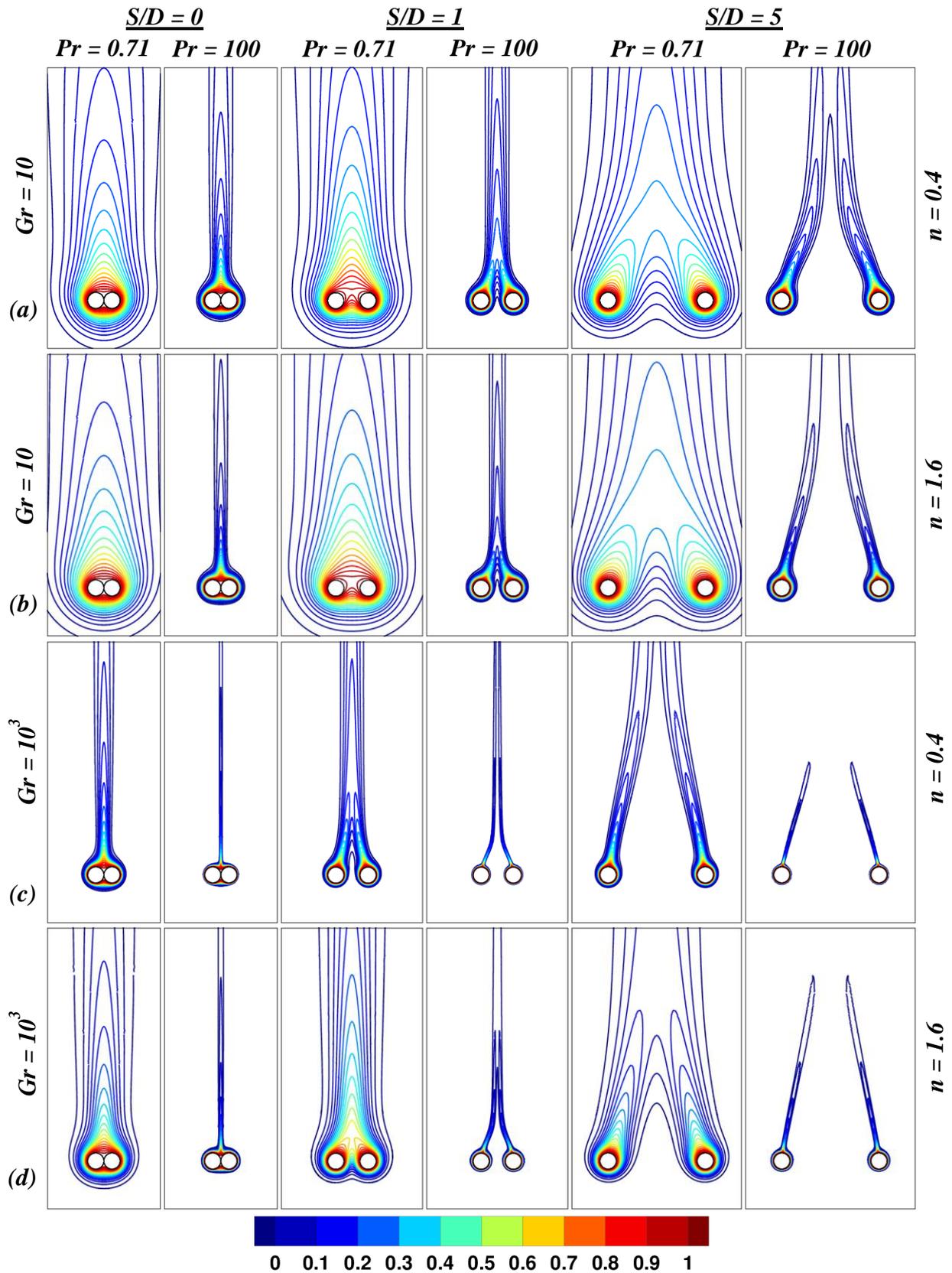

**Figure 7.** Normalized isotherm contours

Finally, it is worthwhile to mention here that at a low value of Gr and Pr (at low Rayleigh number), the mode of heat transfer is primarily by conduction and hence, the effect of the power-law index has negligible influence on the thermal field as shown in Figure 7. In addition, at S/D = 0, the recirculation zones (vortices) have relatively less influence on the thermal field at low and moderate Grashof numbers whereas, at high Grashof number (Gr = $10^5$), it may have a significant influence on heat transfer in some cases, as reported in the literature [11,33]. This effect is also found in the present study and discussed in the next section under local Nusselt number distributions.

### 4.4. Local Nusselt number distributions

The local heat transfer characteristics are elucidated in terms of the distribution of local Nusselt numbers over the surface of the cylinders, which in turn can provide further physical insights into the flow and thermal fields. Owing to the symmetry of the flow and thermal plume about the vertical centreline of the computational domain, the representative results of the local Nusselt number distribution are shown in Figure 8, Figure 9, and Figure 10 only for the right cylinder (since the left cylinder would have an exact mirror image effect) annotating the combined effects over the range of pertinent conditions, horizontal spacing (S/D = 0, 1, and 5), Grashof number ($10 \leq Gr \leq 10^5$), Prandtl number ($0.71 \leq Pr \leq 100$), and Power-law index (n = 0.4, 1 and 1.6; manifesting shear-thinning, Newtonian and shear-thickening fluid behavior, respectively).

For a Newtonian fluid, the local value of the Nusselt number is solely governed by the temperature gradient normal to the surface of the cylinders since the viscosity of the fluid is constant throughout the computational domain. Conversely, for a non-Newtonian Power-law fluid, in addition to the temperature gradient, the local value of the Nusselt number is also governed by the local apparent viscosity (which changes progressively with shear-rate) of the fluid on the surface of the cylinders where the local values of Gr and Pr are expected to be different from those at the far field conditions due to the appearance of the power-law index in the definition of both Gr and Pr as given by Eq.(16) and Eq.(17), respectively. In addition, the temperature gradient is further influenced by the relative local magnitude of the two causes such as the temperature difference between the cylinder surface and the neighboring fluid along with the velocity gradient at the surface of the cylinders. It can be seen from Figure 8, Figure 9, and Figure 10 that at a low value of Grashof number or Prandtl number, the local Nusselt number changes very moderately with the circumferential angle (θ) over the surface of the cylinders, owing to the poor advection or dominating effect of conduction. Irrespective of the power-law index and horizontal spacing, the value of the local Nusselt number increases progressively with increase in Gr or Pr or both, attributed to the thinning of the thermal boundary layers, in turn, increases the temperature gradient which was already anticipated from the isotherm contours in Figure 7 in the preceding section. Overall, the value of the local Nusselt numbers are found to be higher in shear-thinning fluids than that of shear-thickening fluids. The overall trends and the key findings from this analysis are summarized as follows:

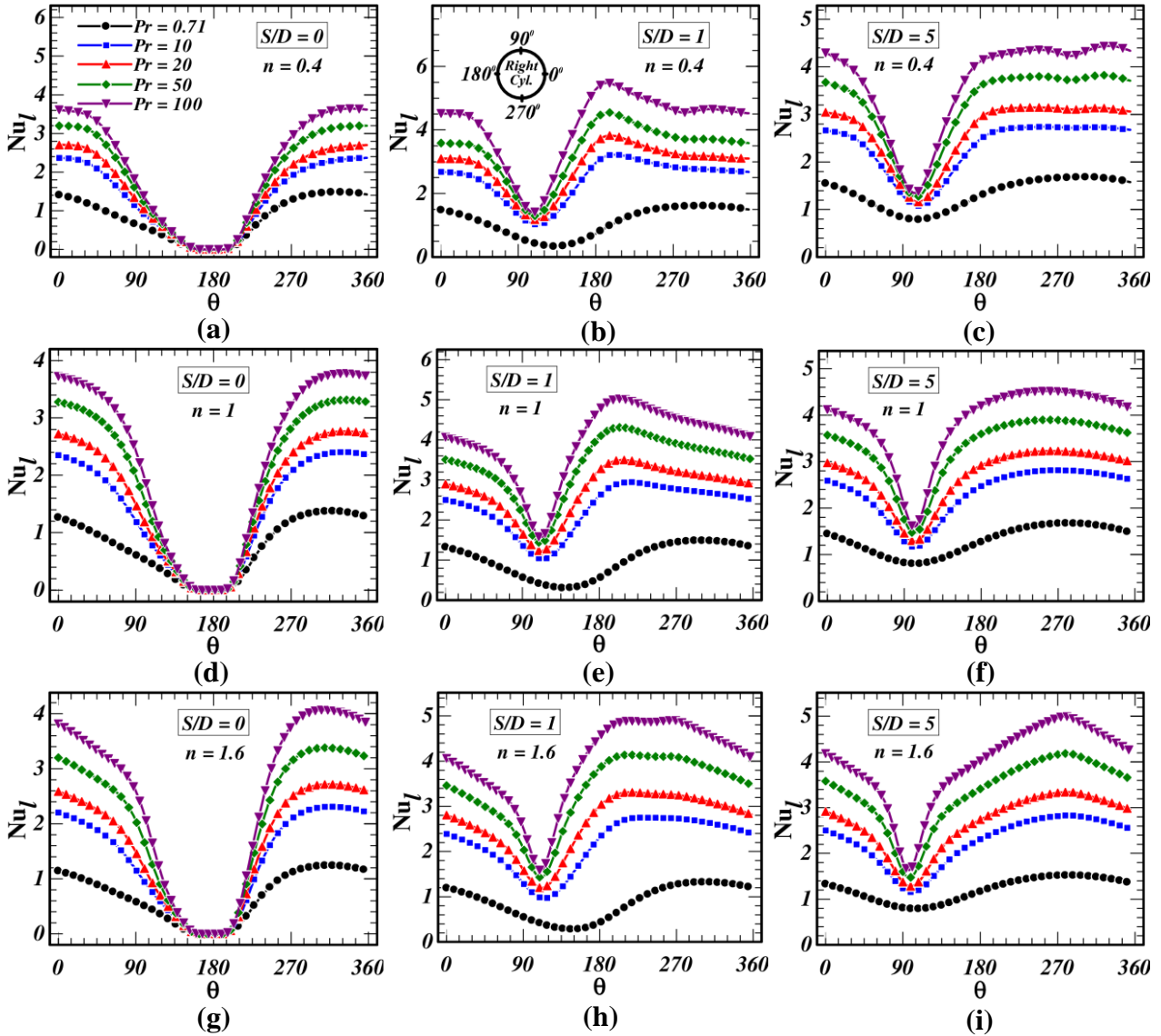

**Figure 8.** Local Nusselt number distribution along the surface of the cylinder as a function of Prandtl number, power-law index and S/D at Gr = 10

The effect of S/D has significant influence on the local Nusselt numbers albeit they follow somehow the same pattern for all S/D except for S/D = 0 (attached cylinders) where the variation of local Nusselt numbers is completely different from those of others due to the compact geometrical structure and strong interaction of the thermal plumes. For S/D = 0, the maximum value of local Nusselt number is seen to be in a region beyond the front stagnation point ($\theta = 270^0$), approximately $\theta = 290^0\text{-}340^0$ depending on Gr, Pr and n. In contrast, for high S/D, the maximum value of local Nusselt number is found to be at the front stagnation point ($\theta = 270^0$), indistinguishable from that of a single horizontal cylinder [7]. With decrease in S/D, the location of maximum heat transfer progressively moves downstream on the symmetry side (moves from θ =270⁰ to 180⁰) and corresponding to the optimal spacing (due to the formation of a chimney effect), the maximum local Nusselt number is found to be higher. Furthermore, the location of maximum local Nusselt number moves downstream from θ = 270⁰ to 180⁰ with a decrease in the power-law index from shear-thickening to shear-thinning.

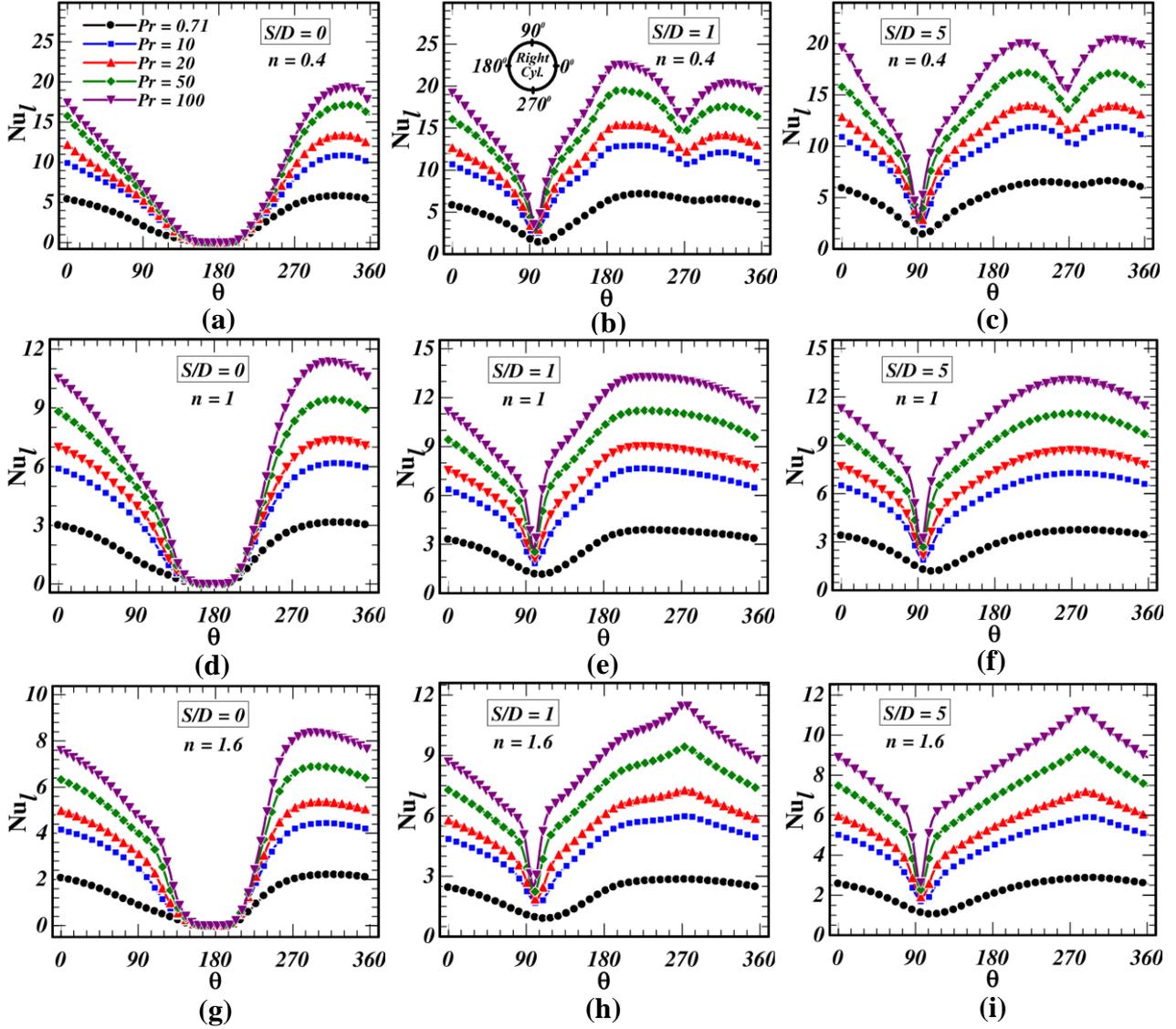

**Figure 9.** Local Nusselt number distribution along the surface of the cylinder as a function of Prandtl number, power-law index and S/D at Gr = $10^3$

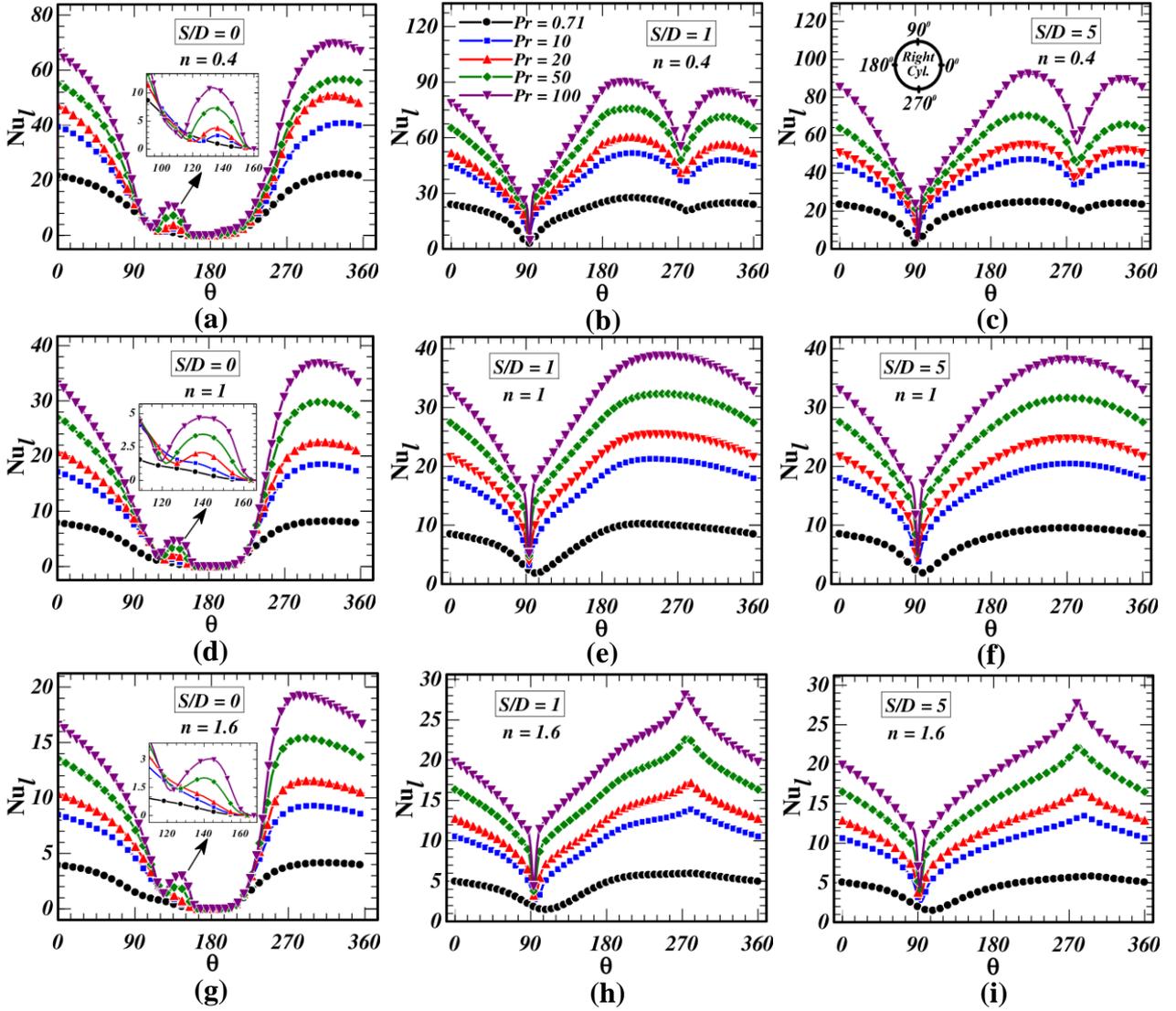

**Figure 10.** Local Nusselt number distribution along the surface of the cylinder as a function of Prandtl number, power-law index and S/D at $Gr = 10^5$

For S/D = 0, the local Nusselt number monotonically decreases over the cylinder surface to a zero value from the starting point ($\theta = 0^0$ or $360^0$) to $\theta = 180^0$ due to gradual decrease in temperature gradient over the surface and the Nusselt number remains zero in the region from $\theta = 150^0\text{-}210^0$ (depending on Gr, Pr and n) called as zero value region. This may be attributed to the merging of two thermal plumes, leading to the formation of flow separations and vortex formations near the contact point, such that the temperature gradient reduces significantly at this region. Beyond this region, the value of local Nusselt number increases monotonically with $\theta$, due to increase in thermal gradient with a favorable pressure gradient over the surface and attains the maximum value at approximately $\theta = 290^0\text{-}340^0$. In addition, the effect of temperature inversion can be clearly seen in Figure 10 (a), (d), and (g) at high Grashof number of $10^5$. Due to a temperature inversion, the local value of Nusselt number suddenly changes from its regular decreasing pattern to an increasing and then decreasing pattern with a local maximum. It can also be noticed that the effect of temperature inversion is prominent at high Grashof number (Gr = $10^5$) and Prandtl number of 10 onwards, (Pr ≥ 10) for shear-thinning fluids, Pr ≥ 20 for Newtonian fluids and Pr ≥ 50 for shear-thickening fluids.

The effective viscosity of a shear-thinning fluid (n < 1) decreases with increase in shear-rate whereas for a shear-thickening fluid (n > 1), the effective viscosity increases with increase in shear-rate over the cylinders. For S/D > 0, the shear rate is expected to be minimum at the front stagnation point (θ = $270^0$) and increases towards the downstream hence, the local heat transfer at the front stagnation point (θ = $270^0$) is found to be lower in shear-thinning fluids and higher in shear-thickening fluids with reference to that of a Newtonian fluid. In addition, owing to the stronger momentum of the flow in the passage between the cylinders and gradual decrease in the velocity gradient over the cylinders, the local Nusselt number is found to be minimum at the back stagnation point (θ = $90^0$) as shown in Figure 8, Figure 9, and Figure 10.

Furthermore, the rate of decrease or increase in the local Nusselt number is more prominent in shear-thickening fluid compared to that of a shear-thinning fluid due to a decrease in the effective viscosity of a shear-thickening fluid with an increase in the shear-rate. Apparently, in the present study, the local heat transfer characteristics are highly distinguishable from that of a single horizontal cylinder [7]. Hence, it is fair to postulate from these conjectures that the global characteristics would also predict differently from those of a single cylinder, which is discussed in the next section as average surface Nusselt number.

## *4.5. Effect on average Nusselt number*

It is customary to figure out the global heat transfer characteristics in terms of the average surface Nusselt number as expressed in Eq. (18), which has great importance in process engineering and industrial design calculations. The dependence of average surface Nusselt number on Grashof number, Prandtl number, and power-law index are shown in Figure 11. As anticipated from the preceding discussions, irrespective of the power-law index, the average Nusselt number increases with increase in Gr and Pr due to progressive thinning of the thermal boundary layer. Conversely, with an increase in the power-law index from shear-thinning (n < 1) to shear-thickening (n > 1), the value of average Nusselt number progressively decreases except for a particular case at Gr = 10. At a very low value of Gr, owing to the slower reduction in apparent viscosity over the surface of the cylinders and a little higher viscosity at beneath the cylinders for shear-thinning fluids, the average Nusselt number initially increases from n=0.4 to 0.6, beyond which it decreases marginally or remains almost constant as shown in Figure 11. This effect is more prominent at high Pr and even more significant at S/D = 0 (attached cylinder) where the average Nusselt number progressively increases with the power-law index as shown in Figure 11 (a). Furthermore, at low Grashof number (Gr = 10), conductive heat transfer dominates over advection hence, the variation of average Nusselt number with the power-law index is seen to be moderate. Overall, the average Nusselt number shows a positive dependence on both Grashof and Prandtl numbers, whereas it shows an adverse dependence on Power-law index.

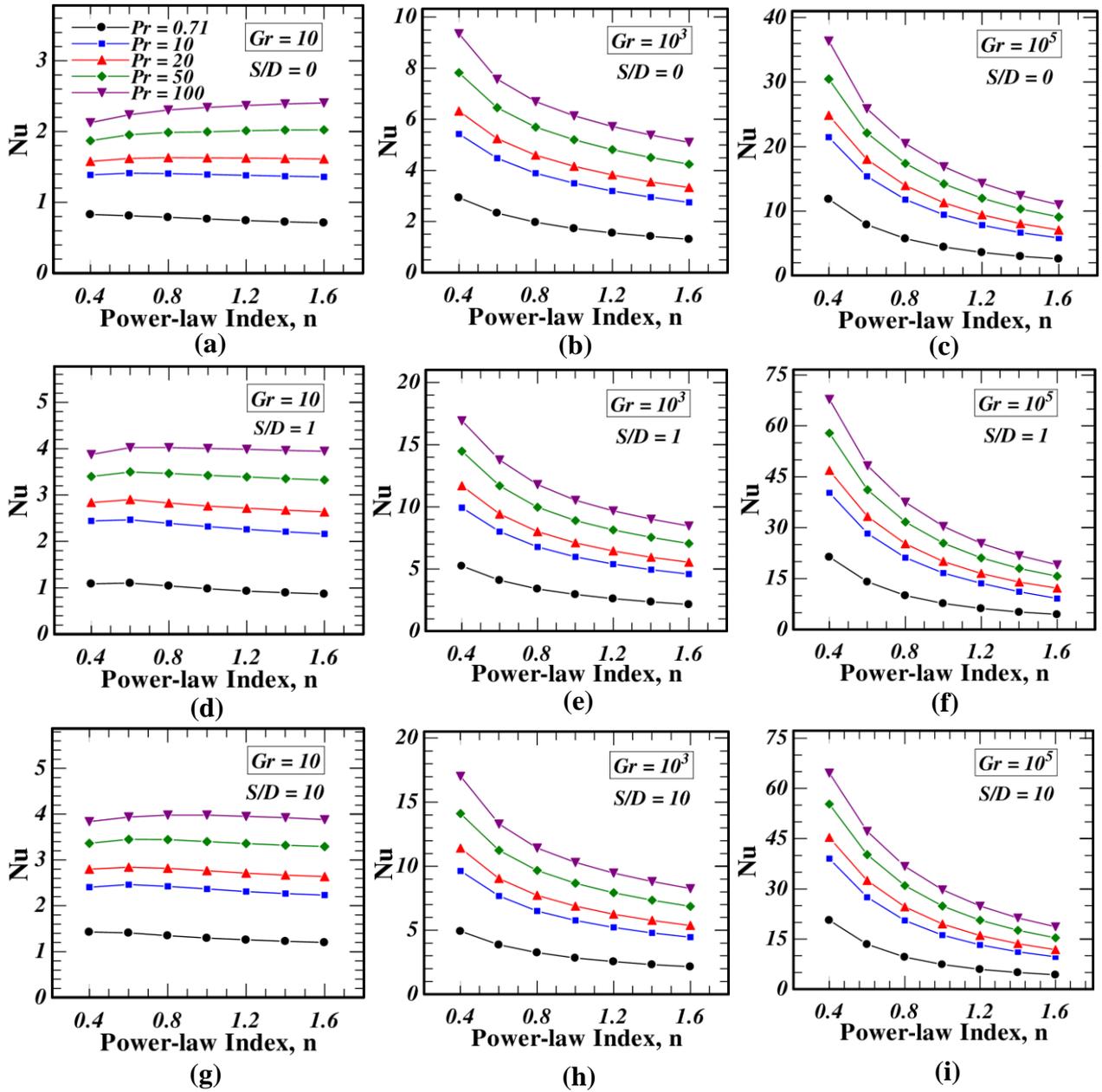

**Figure 11.** Dependence of average Nusselt number on Grashof number, Prandtl number, power-law index, and S/D

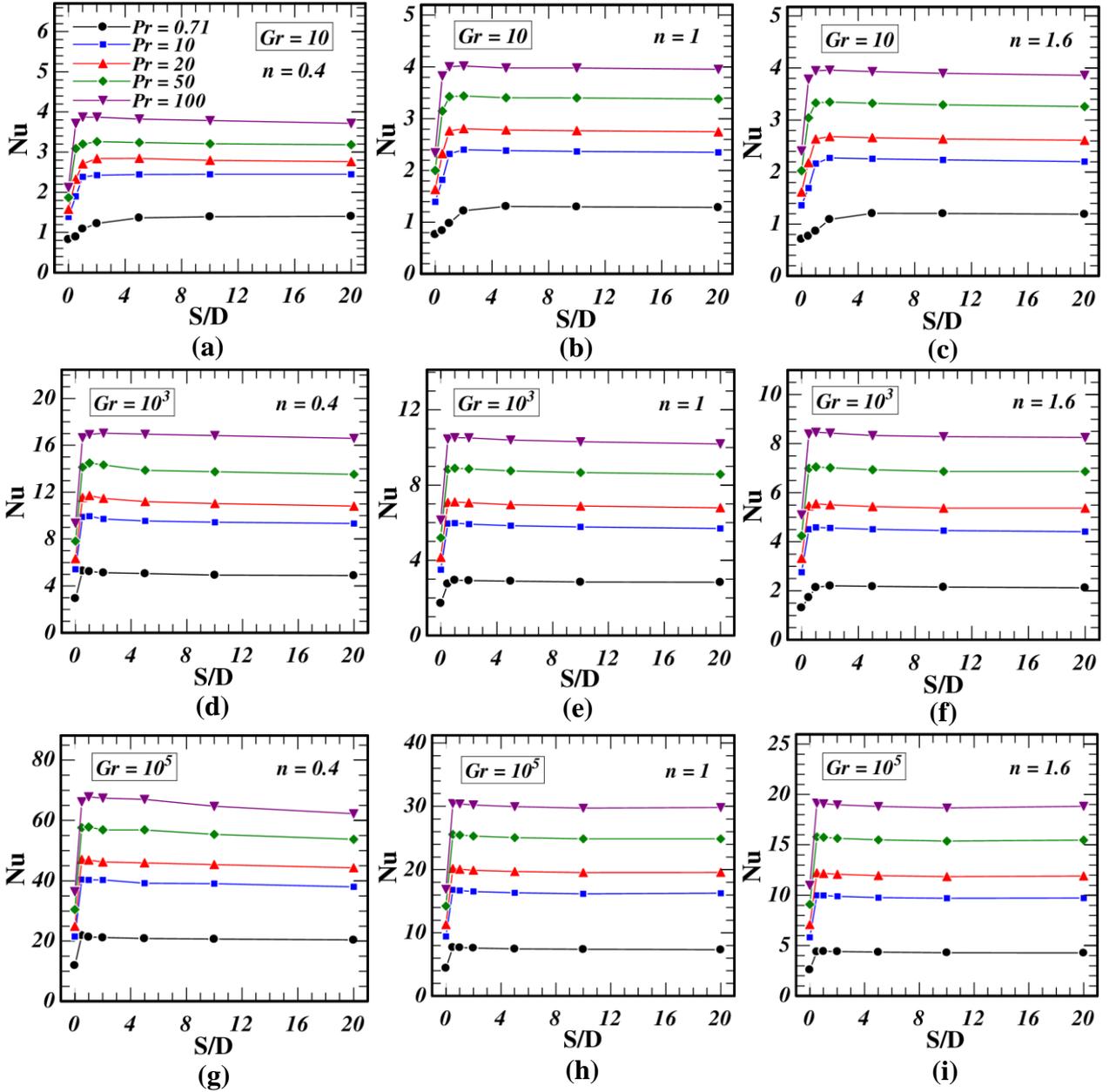

**Figure 12.** Effect of S/D on average Nusselt number as a function of Grashof number, Prandtl number and power-law index

The effect of S/D on the average surface Nusselt number is shown in Figure 12. As postulated from the preceding discussions of thermal field[§4.3] and local Nu distributions[§4.4], It can be clearly seen that owing to the development of a narrow passage between the two cylinders (chimney effect), the average Nusselt number significantly increases to the optimal value when the horizontal spacing increases from S/D = 0 to the optimal spacing. With further increase in the horizontal spacing, the average Nusselt number is found to decrease marginally. Furthermore, the optimal horizontal spacing is found to be in the range S/D = 0.5 to 2 depending on Gr, Pr, and n.

*4.6. Local drag coefficient distributions*

The local distribution of drag coefficients over the surface of the cylinders are solely associated with the velocity gradient at the surface, which can illustrate many physical insights into the flow field. The local distribution of the two components of the total drag coefficient as expressed in Eq.(20) such as pressure drag coefficient, Eq.(22) and friction drag coefficient, Eq.(24) are plotted separately as a function of S/D, Gr, Pr, and n. For the sake of brevity, the local distributions are not shown here for $Gr = 10^5$, which would show the same pattern as $Gr = 10$ and $10^3$.

4.6.1. *Distribution of local pressure drag coefficients:*

Figure 13 shows the distribution of local pressure drag coefficients with the circumferential angle (θ) over the surface of the cylinder for two extreme values of the power-law index (n = 0.4 and 1.6; exhibiting a shear-thinning and shear-thickening behaviors, respectively) and for two horizontal spacing cases such as; no spacing (S/D = 0) & with spacing (S/D = 1). The local pressure drag is attributed to the pressure difference between the local pressure on the cylinder surface and the pressure of the quiescent power-law fluid as mentioned in Eq. (22). For S/D = 0, the pressure coefficient is found to be maximum at the front stagnation point or below the contact point ($\theta = 180^0$) and minimum at the back stagnation point or above the contact point ($\theta = 180^0$). For S/D > 0, the front and back stagnation points have been shifted towards the bottom ($\theta = 270^0$) and top ($\theta = 90^0$) of the cylinders, respectively. The pressure coefficient gradually decreases from its front stagnation point to the back stagnation point, due to the increase in velocity of the flow or decrease in the favorable pressure gradient over the surface of the cylinders. At low Prandtl number, the viscosity of the fluid is low hence, the rate of decrease or increase in pressure coefficient is more prominent for the low value of Pr and this rate gradually decreases with increase in Pr. With increase in shear-rate, the apparent viscosity of a shear-thinning fluid increases hence, the pressure coefficient is found to be maximum for a shear-thinning fluid and minimum for a shear-thickening fluid. Furthermore, with an increase in Grashof number, the momentum of the flow significantly increases hence, the range of pressure coefficient decreases remarkably with increase in Grashof number as shown in Figure 13. The high-pressure zone gradually shrinks and the high-pressure value progressively decreases with increase in S/D as postulated from the pressure contours[§4.2]. Hence, the pressure coefficient is found to be higher for S/D =0 (no spacing) and lower for S/D =1 (with spacing) as shown in Figure 13.

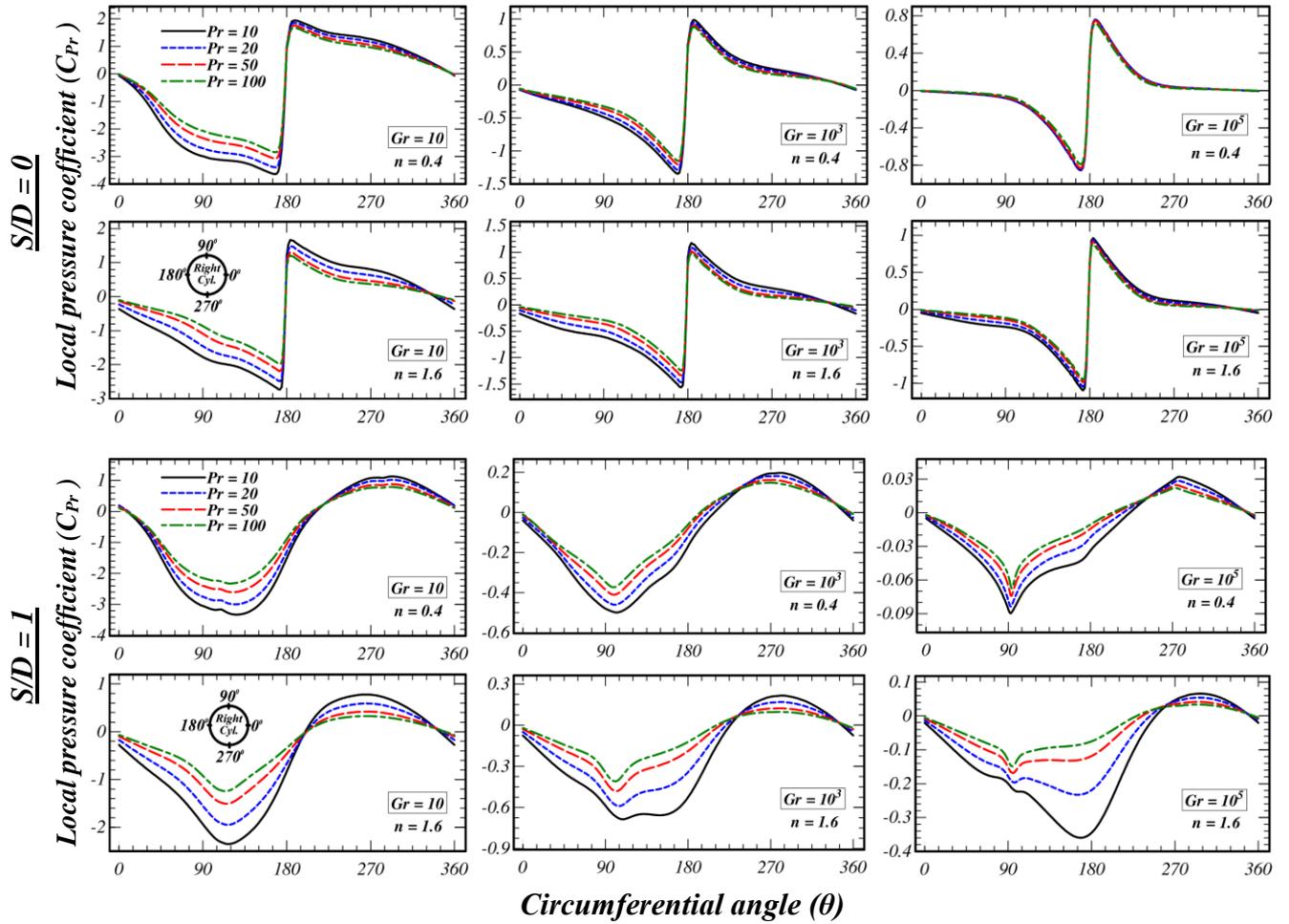

**Figure 13.** Local Pressure coefficient distribution along the surface of the cylinder as a function of Grashof number, Prandtl number, power-law index, and S/D

*4.6.2. Distribution of local skin-friction drag coefficient:*

The local friction drag is associated with the local velocity gradient at the surface of the cylinders as expressed in Eq. (24), which varies along the surface of the cylinders. The distribution of the local skin-friction drag coefficients with the circumferential angle ($\theta$) over the surface of the cylinders are shown in Figure 14 for two extreme values of the power-law index (n = 0.4 and 1.6) and for two horizontal spacing cases such as; no spacing (S/D = 0) & with spacing (S/D = 1). For S/D =0 (attached cylinders), the local value of the friction drag coefficient is found to be maximum at around $\theta = 0^0$ where the velocity gradient is expected to be maximum, except for a particular case at Gr = 10 and n = 0.4 where the maximum value is shifted towards downstream around $\theta = 45^0$-$60^0$ due to delay in the development of boundary layer. In contrast, for S/D >0, owing to the development of a chimney effect, the velocity of the flow is expected to be maximum on the symmetry side around $\theta = 180^0$ compared to that of the other side around $\theta = 0^0$. Hence, for S/D =1, the local friction drag coefficient is found to be higher on both horizontal sides of the cylinders albeit it shows a maximum value on the symmetry side (θ = 180⁰). From the location of higher skin friction value, the boundary layer develops in both upstream & downstream and the velocity gradient decreases progressively hence, the friction drag coefficient gradually decreases over the cylinder surface and reaches to a minimum value at the front and back stagnation points as shown in Figure 14.

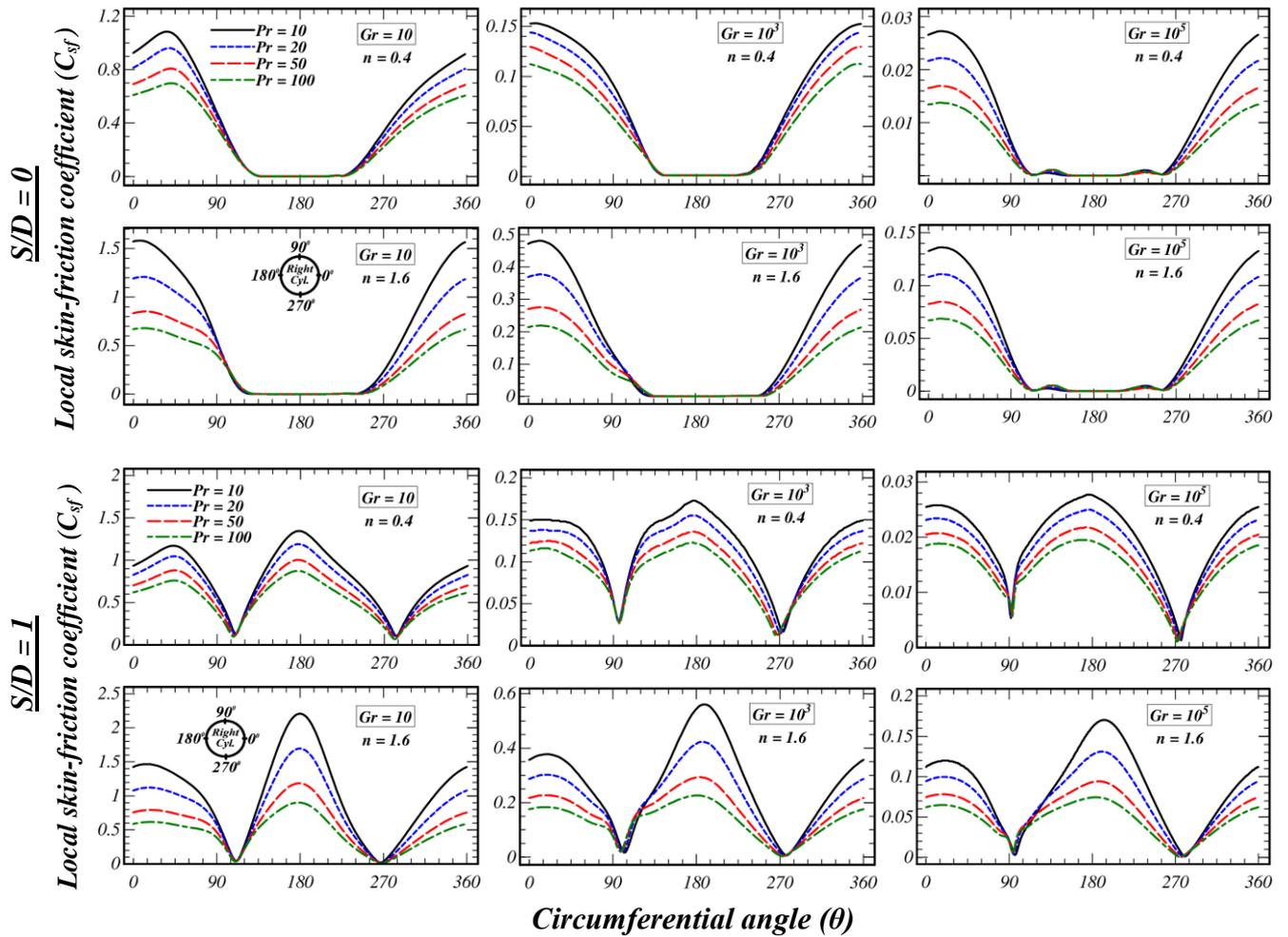

**Figure 14.** Local skin-friction coefficient distribution along the surface of the cylinder as a function of Grashof number, Prandtl number, power-law index, and S/D

For S/D =0, the region of minimum skin friction coefficient increases with increasing value of the Gr, due to increase in the size of the vortices. It can be seen that with increase in the horizontal spacing from S/D =0 to the optimal spacing, the maximum value of the local friction coefficient increases (owing to the chimney effect) which is expected to decrease with further increase in S/D. With increase in Prandtl number, the diffusion of momentum decreases and as a result the local skin friction coefficients monotonically decreases. Furthermore, with an increase in Grashof number, the momentum or velocity of the flow increases but due to the non-dimensional scheme, the skin friction coefficients are found to decrease with increase in Gr as shown in Figure 14.

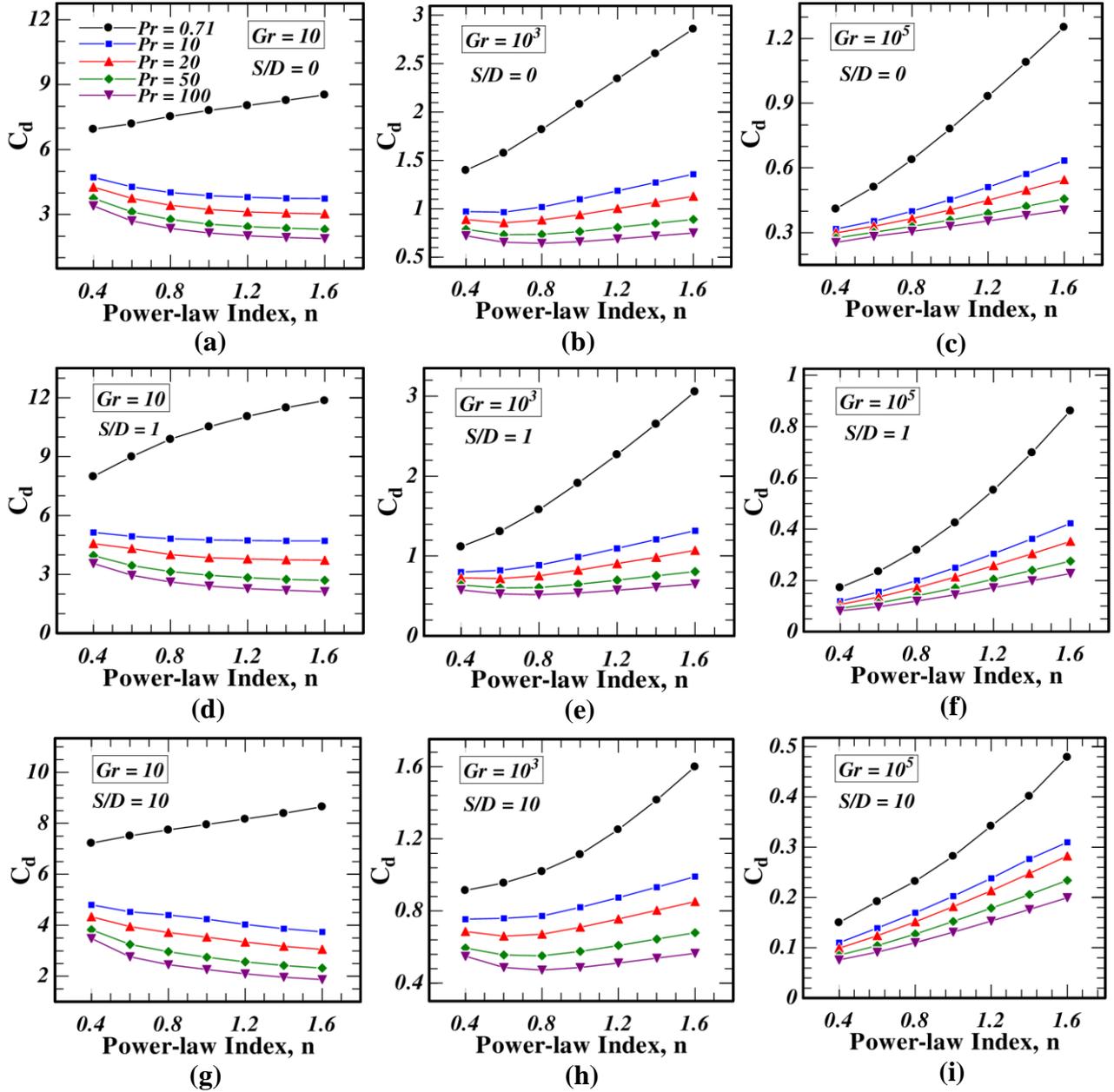

**Figure 15.** Dependence of total drag coefficient on Grashof number, Prandtl number, power-law index, and S/D

## *4.7. Effect on total drag coefficient*

The normal and tangential components of the force exerted by the fluid on the surface of the cylinders give rise to a net buoyancy force in the positive Y-direction, which is usually expressed non-dimensionally in terms of the drag coefficient ($C_d$) as defined in Eq. (20). Figure 15 represents the functional dependence of the total drag coefficient on the Power-law index, Grashof number, Prandtl number, and horizontal spacing. It can be seen that the total drag coefficient decreases with increase in Prandtl number and Grashof number due to increase in the inertia of the flow (increase in the free convection Reynolds number and also partly due to the non-dimensional scheme adopted in the study).

It is customary to mention here that for a non-Newtonian power-law fluid, the inertia and viscous forces are scaled as ~ $u_c^2$ and ~ $u_c^n$, respectively. The strength of the buoyancy is weak (low $u_c$) at a low Grashof number hence, the influence of power-law index on drag coefficient is found to be insignificant at low Grashof number (Gr = 10). The effect of power-law index is more prominent in shear-thickening (n > 1) fluids compared to that in Newtonian (n = 1) and shear-thinning (n < 1) fluids. At low Gr, pressure drag coefficient dominates over the friction drag coefficient and vice-versa. Hence, with increase in the power-law index, the total drag coefficient decreases at a low Grashof number (Gr = 10) and increases at moderate and high Grashof numbers. For Pr <1, the total drag coefficient divulges an anomalous behavior owing to the dominating effect of thermal diffusion over the diffusion of momentum. In addition, the trends of the total drag coefficient as seen in the present study are qualitatively similar to that of a single cylinder [7].

Furthermore, the effect of horizontal spacing plays a significant role on the total drag coefficients. Figure 16 shows the effect of horizontal spacing on the total drag coefficient for some pertinent conditions. At low Gr, owing to the conduction dominating effect, the total drag coefficient increases from S/D =0 to the optimal spacing (around S/D =0.5 to 1) and with further increase in S/D, it progressively decreases. In contrast, at moderate and high Gr, due to the stronger chimney effect, the total drag significantly decreases from no spacing (S/D =0) to the optimal spacing and beyond which it progressively decreases. The effect of S/D is more prominent at low Prandtl number and high power-law index (Newtonian and shear-thickening fluids; n ≥1) as shown in Figure 16.

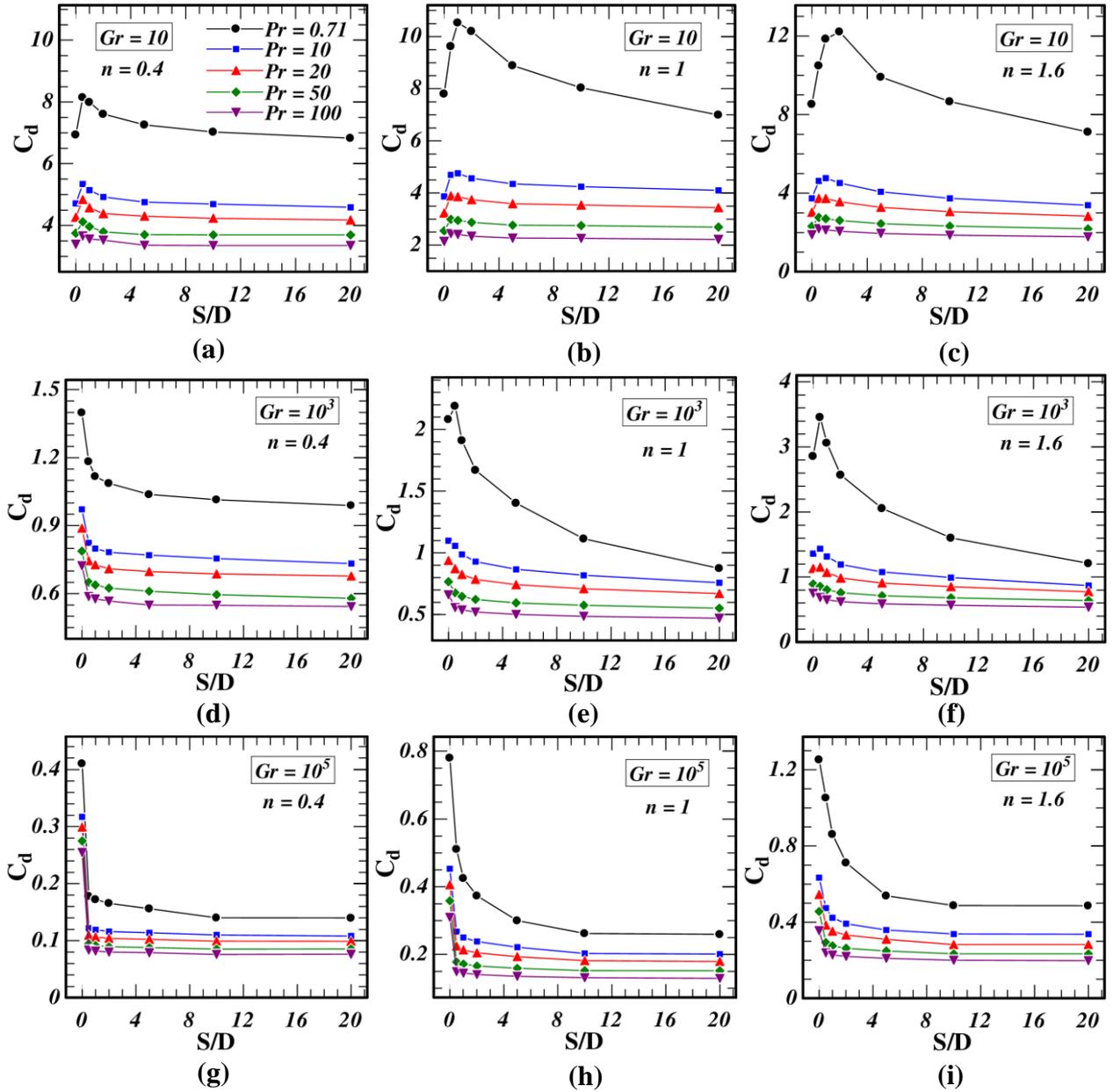

**Figure 16.** Effect of S/D on total drag coefficient as a function of Grashof number, Prandtl number and power-law index

## 5. Correlation for average Nusselt number

The heat transfer correlations are mostly implemented for industrial design calculations, which can also be useful for practicing engineers and academic researchers. Hence, an empirical correlation for the average Nusselt number is developed by using a non-linear regression analysis of the present computational results. In the present study, the average Nusselt number has a functional relationship with horizontal spacing, Grashof number, Prandtl number, and Power-law index hence, the average Nusselt number is correlated in a form as expressed in Eq.(25), which is valid in the following range of conditions: $0 \leq S/D \leq 20$, $10 \leq Gr \leq 10^5$, $0.71 \leq Pr \leq 100$, and $0.2 \leq n \leq 1.8$.

$$\mathrm{Nu} = a\left[\mathrm{Gr}^{\frac{1}{2(n+1)}} \mathrm{Pr}^{\frac{n}{3n+1}}\right]^b \times \left[1 + \frac{1}{(S/D)}\right]^c \quad (25)$$

The correlating constants (a, b, and b) along with the coefficient of the correlations ($R^2$) are shown in Table 4. Figure 17 shows a good agreement between the numerically calculated Nusselt number and the predicted values of Nusselt number from the correlation (Eq. (25)), which shows more than 90% data points within the accuracy range of ±10%.

**Table 4.** Correlation constants with coefficient of correlation ($R^2$) for the average Nusselt number correlations

| *a* | *b* | *c* | *$R^2$* |
|---|---|---|---|
| 0.6842 | 0.8866 | 0.1880 | 0.9904 |

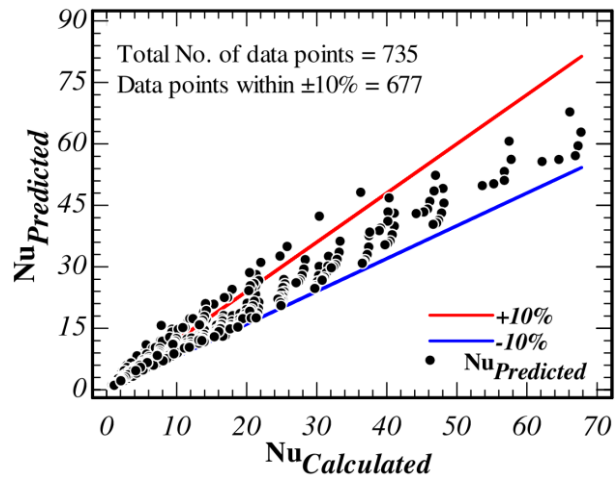

**Figure 17.** Predicted and computed values of average Nusselt number

## 6. Concluding Remarks

In this work, the effect of horizontal spacing between two horizontally aligned circular cylinders on natural convection heat transfer of non-Newtonian Power-law fluids has been investigated numerically over a wide range of pertinent conditions: $10 \leq \mathrm{Gr} \leq 10^5$, $0.71 \leq \mathrm{Pr} \leq 100$, and $0.4 \leq n \leq 1.6$. The present numerical simulations have been conducted for the horizontal spacing spanning in the range $0 \leq S/D \leq 20$ and some of the major findings are enumerated below:

*(a)* The thermal boundary layers are seen to be thinning progressively with decrease in the power-law index or with increase in Grashof number or Prandtl number or both. Hence, an augmentation in heat transfer is anticipated from this behavior.

*(b)* At S/D = 0, two recirculation zones are seen to form on both above and below the contact point of the two attached cylinders and the size of the vortices grows with increase in Grashof number and shrinks with increase in Prandtl number. In contrast, the vortex formations are completely absent for S/D > 0.

*(c)* The local and average values of Nusselt number are found to show a positive dependence on both Grashof number and Prandtl number whereas an adverse dependence is seen on Power-law index.

*(d)* Overall, under identical conditions, shear- thinning fluid (n < 1) enhances the convection heat transfer whereas shear- thickening fluid (n > 1) impedes it with reference to a Newtonian fluid (n = 1).

*(e)* Owing to the development of a chimney effect, the heat transfer rate gradually increases with decrease in S/D and reaches to a maximum value corresponding to the optimal spacing. With a further decrease in S/D, the two thermal plumes interact with each other and as a result, heat transfer rate diminishes.

*(f)* The average Nusselt number is minimum as S/D =0 (no spacing) and found to be 21 to 89% higher at the optimal horizontal spacing depending on the value of Grashof number, Prandtl number, and power-law index.

*(g)* All else being equal, the total drag coefficient shows an adverse dependence on both Grashof and Prandtl number whereas it shows a positive dependence on power-law index at moderate and high Grashof number and due to weak momentum of the buoyancy flow at low Grashof number (Gr=10), the dependence of power-law index is seen to be insignificant.

*(h)* With increase in S/D, the total drag coefficient decreases progressively at moderate and high Gr, whereas at low Gr, it initially increases up to an optimal horizontal spacing and beyond which it gradually decreases.

*(i)* Finally, an empirical correlation for the average Nusselt number is developed as a function of S/D, Gr, Pr, and n which can be useful to process industrial calculations and academic researchers.

**Acknowledgments**

The present research work was carried out in Computational Fluid Dynamics (CFD) laboratory of the Department of Mechanical Engineering at the Indian Institute of Technology Kharagpur, India.

*Nomenclatures*

| | | | |
|---|---|---|---|
| $D$ | diameter of the cylinders (m) | $\hat{n}$ | unit vector |
| $D_\infty$ | diameter of the outer domain (m) | $s$ | surface area of the cylinder (m$^2$) |
| $L$ | characteristic length scale (m) | $u$ | velocity of the flow (m/s) |
| $u_c$ | characteristic reference velocity (m/s) | $U$ | non-dimensional velocity of flow |
| $g$ | acceleration due to gravity (m/s$^2$) | $x, y$ | cartesian coordinates (m) |
| $T$ | temperature of the fluid (K) | $X, Y$ | non-dimensional cartesian coordinates |
| $T_m$ | mean film temperature (K); $T_m = (T_w + T_\infty)/2$ | $i, j$ | index notations |
| $\Delta T$ | temperature difference (K); $\Delta T = T_w - T_\infty$ | | |

*Greek symbols*

| | | | |
|---|---|---|---|
| $p$ | pressure (Pa) | $\beta$ | thermal expansion coefficient (1/K) |
| $P$ | non-dimensional pressure | $\rho$ | density of the fluid (kg/m$^3$) |
| $K$ | thermal conductivity of the fluid (W/m K) | $\varphi$ | normalized temperature; $\varphi = (T - T_\infty)/(T_w - T_\infty)$ |
| $C_p$ | thermal heat capacity of fluid (J/kg K) | | |
| $Gr$ | Grashof number (dimensionless) | $\theta$ | circumferential angle |
| $Pr$ | Prandtl number (dimensionless) | $\eta$ | apparent viscosity (Pa s) |
| $m$ | power-law consistency index (Pa s$^n$) | $\tau$ | deviatoric stress tensor (Pa) |
| $n$ | power-law index (dimensionless) | $\varepsilon$ | strain rate tensor (1/s) |
| $Nu$ | Nusselt number (dimensionless) | $\dot{\gamma}$ | shear rate (1/s) |
| $h_l$ | local heat transfer coefficient (W/m$^2$ K) | | |

*Subscripts*

| | | | |
|---|---|---|---|
| $C_d$ | total drag coefficient | $\infty$ | ambient |
| $C_{pr}$ | pressure drag coefficient | $w$ | wall |
| $C_{sf}$ | skin friction drag coefficient | $l$ | local |
| $I_2$ | second invariant of strain rate tensor (s$^{-2}$) | $\theta$ | local points |